\documentclass[fleqn,usenatbib]{mnras}

\usepackage[T1]{fontenc}

\DeclareRobustCommand{\VAN}[3]{#2}
\let\VANthebibliography\thebibliography
\def\thebibliography{\DeclareRobustCommand{\VAN}[3]{##3}\VANthebibliography}

\usepackage{graphicx}	
\usepackage{amsmath}	
\usepackage{amssymb}	
\usepackage[para,online,flushleft]{threeparttable}
\usepackage{soul} 
\usepackage{multirow,booktabs} 
\usepackage{newtxtext,newtxmath} 

\def\la{\mathrel{\hbox{\rlap{\hbox{\lower4pt\hbox{$\sim$}}}{\raise2pt\hbox{$<$}}}}}
\def\ga{\mathrel{\hbox{\rlap{\hbox{\lower4pt\hbox{$\sim$}}}{\raise2pt\hbox{$>$}}}}}

\title[The disc wind models of the ULX NGC 5408 X-1]{X-ray reverberation models of the disc wind in ultraluminous X-ray source NGC 5408 X-1}

\author[Luangtip et al.]{
W. Luangtip,$^{1,2}$\thanks{E-mail: wasutep@g.swu.ac.th}
P. Chainakun,$^{3,4}$\thanks{E-mail: pchainakun@g.sut.ac.th}
S. Loekkesee,$^{1}$
C. Deesamer,$^{3}$
T. Ngonsamrong$^{3}$ and
\newauthor
T. Sintusiri $^{3}$
\\
$^{1}$Department of Physics, Faculty of Science, Srinakharinwirot University, Bangkok 10110, Thailand\\
$^{2}$National Astronomical Research Institute of Thailand, Chiang Mai 50180, Thailand\\
$^{3}$School of Physics, Institute of Science, Suranaree University of Technology, Nakhon Ratchasima 30000, Thailand\\
$^{4}$Centre of Excellence in High Energy Physics and Astrophysics, Suranaree University of Technology, Nakhon Ratchasima 30000, Thailand\\
}

\date{Accepted XXX. Received YYY; in original form ZZZ}

\pubyear{2020}

\begin{document}
\label{firstpage}
\pagerange{\pageref{firstpage}--\pageref{lastpage}}
\maketitle

\begin{abstract}
Majority of ultraluminous X-ray sources (ULXs) are believed to be super-Eddington objects, providing a nearby prototype for studying an accretion in super-critical regime. In this work, we present the study of time-lag spectra of the ULX NGC 5408 X-1 using a reverberation mapping technique. The time-lag data were binned using two different methods: time averaged-based and luminosity-based spectral bins. These spectra were fitted using two proposed geometric models: single and multiple photon scattering models. While both models similarly assume that a fraction of hard photons emitted from inner accretion disc could be down-scattered with the super-Eddington outflowing wind becoming lagged, soft photons, they are different by the number that the hard photons scattering with the wind: i.e. single vs multiple times. In case of averaged spectrum, both models consistently constrained the mass of ULX in the range of ~$\sim$80-500 M$_{\rm \odot}$. However, for the modelling results from the luminosity based spectra, the confidence interval of the BH mass is significantly improved and is constrained to the range of ~$\sim$75-90 M$_{\rm \odot}$. In addition, the models suggest that the wind geometry is extended in which the photons could down-scatter with the wind at the distance of $\sim$10$^{4}$ -- 10$^{6}$ $r_{\rm g}$. The results also suggest the variability of the lag spectra as a function of ULX luminosity, but the clear trend of changing accretion disc geometry with the spectral variability is not observed.

\end{abstract}

\begin{keywords}
accretion, accretion discs -- stars: black holes -- stars: winds, outflows -- X-rays: binaries -- X-rays: individual: NGC 5408 X-1
\end{keywords}



\section{Introduction}

Since the first detection of ultraluminous X-ray sources (ULXs) by \textit{Einstein Observatory} \citep{long1981,fabbiano1989} , they have become an object of interest due to their relatively high luminosity: X-ray luminosity ($L_{\rm X}$) $\ga$ 10$^{39}$ erg s$^{-1}$ exceeding of the Eddington luminosity for a 10 M$_{\rm \odot}$ black hole (BH; see \citealt{kaaret2017} for a recent review). By definition, the ULXs are extragalactic, off-nuclear point like sources, so that they are not supermassive BHs (SMBHs) located at a galactic centre. Indeed, the ULX population is heterogeneous. While some ULXs could be a promising candidate for intermediate mass BHs (IMBHs) accreting matter at sub-Eddington rate, e.g. ESO 243-49 HLX-1 \citep{farrell2009} and M82 X-1 \citep{pasham2014}, it has been thought that the majority of them are stellar mass objects accreting at super-Eddington rate such as stellar mass BHs (sMBHs) or even neutron stars (NSs). In fact, the observational evidences for the detection of neutron star ULXs (NS-ULXs) have emerged increasingly during the past recent years (e.g. \citealt{bachetti2014,furst2016,israel2017a,israel2017b,carpano2018,sathyaprakash2019,rodriguez2020}) -- especially through the detection of pulsating signal -- leading to the assumption that most ULXs might be powered by NS. This would confirm that the ULXs are accreting matter at super-critical rate, perhaps, up to 500 times of Eddington rate (e.g. \citealt{israel2017a}), providing the opportunity to study the accretion at this extreme rate from nearby galaxies.

Previous studies of ULX has often focused on an X-ray spectral analysis since it actually required lower telescope observational time, comparing to the timing analysis; a lot of ULX properties have been revealed. Nowadays, it is accepted that the X-ray spectra of bright ULXs ($L_{\rm X}$ $\ga$ 3 $\times$ 10$^{39}$ erg s$^{-1}$) would compose of two components: the high energy curvature and the soft excess (e.g. \citealt{gladstone2009,sutton2013}). It has been believed that the hard component is the emission from the hot, inner part of the disc while the soft component could be interpreted as the thermal emission associated with an outflowing wind launched from the outer disc (e.g. \citealt{middleton2015}). Recently, the direct detection of the putative outflowing winds have been reported in several ULXs (e.g. \citealt{pinto2016,pinto2017,pinto2021}), confirming the existence of the outflow. However, the details of wind geometry and accretion mechanism are still ambiguous. To understand more about these, the timing analysis would be helpful. In fact, this technique has been applied to study compact object across mass regime: from X-ray binaries to AGN (e.g. \citealt{middleton2007,uttley2011,alston2020,Chainakun2021}). For ULXs, the number of timing studies have increased progressively, especially since the first detection of the NS-ULX, and have revealed much information about the nature and properties of ULXs, in addition to that is obtained from the spectral studies (see \citealt{kaaret2017} and references therein).

X-ray reverberation mapping is one of timing analysis techniques widely used to study the accretion structure and geometry, as well as mass of central compact objects. This technique is based on the measurement of the arrival time delays between the reflection and the continuum photons \citep{uttley2014, Cackett2021}. Since the reflection photons travel an additional distance to the disc before being observed, we expect to see the variations in the reflection dominated energy band lagging behind those of the continuum dominated band. The amplitude of the lags (referred to as reverberation lags) depends on the geometry of the corona and the disc, hence allows us to probe the exact geometry of the central regions. The X-ray reverberation technique was firstly used to study AGN and later was also applied to study black hole binaries (BHBs; see \citealt{uttley2014} and references therein) so that it become one of canonical technique for estimating black hole masses. However, for ULXs, only few sources have been found to exhibit an X-ray reverberation signal (see e.g. \citealt{demaco2013,garcia2015,pinto2017,kara2020}). This is probably because the exposure time of the available observations of ULXs are mostly short and, indeed, the sources themself are likely to have low count rate and low variability. Here, we select to analyse the data of NGC 5408 X-1 which is one of the ULXs that showed well characterised time lags that were investigated with different scenarios using different models of the reflection, but exact geometry of the system still has been under debate.

NGC 5408 X-1 is a nearby ULX ($D$ $\sim$ 4.8 Mpc) located at $\approx$20 arcsec off from the galactic centre of the host galaxy, and the having the peak X-ray luminosity of $L_{\rm X}$ $\sim$ 10$^{40}$ erg s$^{-1}$ \citep{garcia2015}.  It exhibited a variability on various timescales from a few ten seconds to a few months making the source good candidate for a timing study \citep{strohmayer2009b, Heil2010, demaco2013,garcia2015}. In fact, this is one of a couple of sources that have been claimed to detect the reverberation lag (soft lag) as well as the associated narrow band feature, i.e. QPO, from its X-ray data, similar to that is observed in BHBs; this leaded to the early conclusion that the ULX might harbour an IMBH \citep{strohmayer2009a,demaco2013}.  However, the later study of this ULX by \citet{garcia2015} have argued that the origin of the lag might be different from that of the BHBs in which the traditional BHB lag is due to the reflection from the disc. They have alternatively explained that the lag could happen as the hard photons pass through and scatter in an intervening medium, or in the expanded, optically thin, outflowing wind. Subsequently, the detection of an outflowing wind from this ULX has been reported by \citet{pinto2016}, confirming the existence of the outflow and implying that the source could be super-Eddington, stellar mass object.

In this work, we analyse the {\it XMM-Newton} archival data of the ULX NGC 5408 X-1 using reverberation mapping technique in order to study the geometry of the ULX accretion disc and the outflowing wind. For the first time, the time-lag spectra are created as a function of the ULX luminosity and modeled; this is to examine the evolution and detail of the accretion geometry, as well as the mass of compact object powering the ULX. The paper is laid out as follows. In Section~\ref{sec:observations}, we explain how we choose and reduce the X-ray data, and how the ULX time-lag spectra are created. The proposed theoretical models used to fit the time-lag data are described in Section~\ref{sec:models}. Then, the details of the time-lag spectral analysis and results are presented in Section~\ref{sec:Results} and, subsequently discussed in Section~\ref{sec:discussion}. We finally summarise our finding in Section~\ref{sec:conclusion}.

\section{Observations and data reduction} \label{sec:observations}

The observational data of NGC 5408 X-1 analysed in this work were obtained from the {\it XMM-Newton} telescrope \citep{Jansen2001} and were available in the {\it XMM-Newton} Science Archive.\footnote{\url{https://nxsa.esac.esa.int}} We selected the observations from ones which the time lag, especially the soft lag, was well identified in previous studies (i.e. \citealt{demaco2013,garcia2015}) and the duration of observations was longer than 100 ks; the latter is to ensure that the quality of obtained time-lag data is sufficient for modelling. All observations analysed in this work are tabulated in Table~\ref{tab:observations}. We performed the data reduction in the standard way using {\sc xmm-sas} package together with the latest calibration files.\footnote{see {\it XMM-Newton} ABC Guide: \url{https://heasarc.gsfc.nasa.gov/docs/xmm/abc/}} We also inspected and manually removed the observing durations which were highly affected by background flaring events from the observational data. In brief, for each observation, we created single event, high energy light curves in 10 -- 12 keV for both pn and MOS detector data. We then visually inspected the periods that were highly affected by the background flaring events, and determined the threshold count rates which yielded the quiescent background intervals. While the threshold count rates could vary from observation to observation, overall, we found that they were in the range of 0.3 -- 0.6 cnt s$^{-1}$ and 0.1 -- 0.4 cnt s$^{-1}$ for pn and MOS detectors, respectively. After the background flaring events were removed, the total, useful exposure time obtained from all detectors of each observation is shown in column 3 of Table~\ref{tab:observations}.

\begin{center}
\begin{table}
\caption{\label{tab:observations} {\it XMM-Newton} data of NGC 5408 X-1 analysed in this work.}
\centering
\begin{threeparttable}
\begin{tabular}{lcccc}
\hline
Obs. ID$^{a}$ & Obs. date$^{b}$ & Exp.$^{c}$ & Rate$^{d}$& Bin$^{e}$\\
& & (ks) & (counts s$^{-1}$) & \\
\hline
0302900101  &2006-01-13  & 316 &$0.940 \pm 0.004$ & Medium\\
0500750101  &2008-01-13  & 198 &$0.885 \pm 0.004$ & Low \\
0653380201  &2010-07-17  & 218 &$1.029 \pm 0.004$ & High \\
0653380301  &2010-07-19  & 337 &$1.021 \pm 0.003$ & High\\
0653380401  &2011-01-26  & 314 &$0.974 \pm 0.004$ & Medium \\
0653380501  &2011-01-28  & 320 &$0.925 \pm 0.003$ & Medium \\
\hline
\end{tabular}
\begin{tablenotes}
{Notes.} $^{a}$The observational ID of the data and $^{b}$its corresponding observing date. $^{c}$Total exposure time of the data obtained from pn, MOS1 and MOS2 detectors after removing the background flaring events. $^{d}$The pn detector's count rate in full band (0.3-10 keV). $^{e}$The group that the data belong to for the analysis classified by the count rate shown in column 4 (see text for detail).
\end{tablenotes}
\end{threeparttable}
\end{table}
\end{center}

The timing data were extracted using the IDL routines {\sc xmm extract}.\footnote{The codes were developed by Simon Vaughan, available at \newline \url{https://www.star.le.ac.uk/sav2/idl.html}} In brief, the light curves were created with the time bin resolution of 10.4~s from the data with having $\rm PATTERN\le4$ and $\rm FLAG=0$ for pn detector and $\rm PATTERN\le12$ for MOS1 and MOS2 detectors. The source extraction region was the circular area centred at the ULX position with the radius of 40 arcsec while the associated background was obtained from the source-free, rectangular region of 225 arcsec $\times$ 125 arcsec. These criteria were applied to create the light curves in two different energy bands -- soft (0.3--1 keV) and hard (1--7 keV) bands -- for all pn, MOS1 and MOS2 data. Note that, here, the photons in the soft band here are likely associated with the reflection from the outflowing wind, as the presence has been reported by \cite{pinto2016}, while those of the hard band are mostly from the inner-disc emission; this should allow us to probe the timing properties, as well as the interaction between the photons, of these two components (see e.g. \citealt{demaco2013}). In other words, the 0.3--1 keV and 1--7 keV bands are representative of the reverberation dominated and direct-continuum dominated bands, respectively, as were previously used by \cite{Heil2010} and \cite{demaco2013}. Moreover, this assumption has been established well by the results from spectral analyses in which the empirical boundary between the soft and hard energy bands is $\sim$1 keV (e.g. \citealt{gladstone2009,sutton2013}). In order to create the time-lag spectrum, all lightcurves in each energy band were divided into individual segments of 1 ks, then converted into cross-spectra and averaged across to obtain the averaged cross-spectrum. The cross-spectra obtained from the two energy bands were then rebinned logarithmically -- in which the bin width of the next, higher frequency bin was increased by a factor of 1.3, relative to that of its adjacent lower frequency bin  -- to improve the signal-to-noise ratio and, finally, derived into the phase and time delay spectrum.

Furthermore, to probe the timing properties of the ULX as a function of accretion rate and luminosity, we also binned the data into three sub-groups according to the observed count rate in full (0.3 -- 10 keV) band obtained from the pn camera: low  (count rate $<$ 0.9 cnt s$^{-1}$), medium  (0.9 cnt s$^{-1}$ $<$ count rate $<$ 1.0 cnt s$^{-1}$) and high (count rate $>$ 1.0 cnt s$^{-1}$) luminosity bins (see column 4 and 5 of Table~\ref{tab:observations}). Then, the time-lag spectrum of each sub-group was created using the same method explained above. All obtained time-lag spectra were then used in the analysis described in Section~\ref{sec:Results}.

\section{Theoretical models} \label{sec:models}

In this work, we have proposed theoretical models to explain the observed time-lag spectra of the ULX NGC 5408 X-1. Assuming that the ULX is accreting material above the Eddington rate, the outflowing wind is expected to be launched (see e.g. \citealt{poutanen2007,pinto2016}). Thus, the geometry of our models is based on this assumption in which the optically thick outflow would be launch from the wind launching radius $r_{\rm w}$, which could create accretion disc with a central opening funnel appearing to the photons emitted from the inner disc. Note that a model with a wind geometry was also used and investigated by \cite{middleton2015}. If an observer observes the ULX nearly face-on or at low inclination angle, a fraction of inner disc (hard) photons could be seen directly by the observer in the hard energy band while some portion of inner disc photons could be down-scattered by the wind becoming the soft photons appearing to the observer. Thus, the soft photon should be delayed by the time that inner disc photons travel until hit the wind surface, comparing to the hard photons that travel directly to an observer. In addition, we have also assumed that the degree of the wind opening angle might be a function of accretion rate in which the angle is smaller as the ULX accretion rate increases \citep[e.g.][]{luangtip2016,pintore2021}. In fact, this would imply that, at higher accretion rate, the inner disc photons could scatter with the wind surface multiple times before escape to the observer. Thus, in this section, we have built both single and multiple scattering models to explain the observed time lag of the photons at lower and relatively higher accretion rates, respectively.

\subsection{Single scattering}
\label{sec:sgscattering} 

\begin{figure*}
	\includegraphics[width=15.0cm]{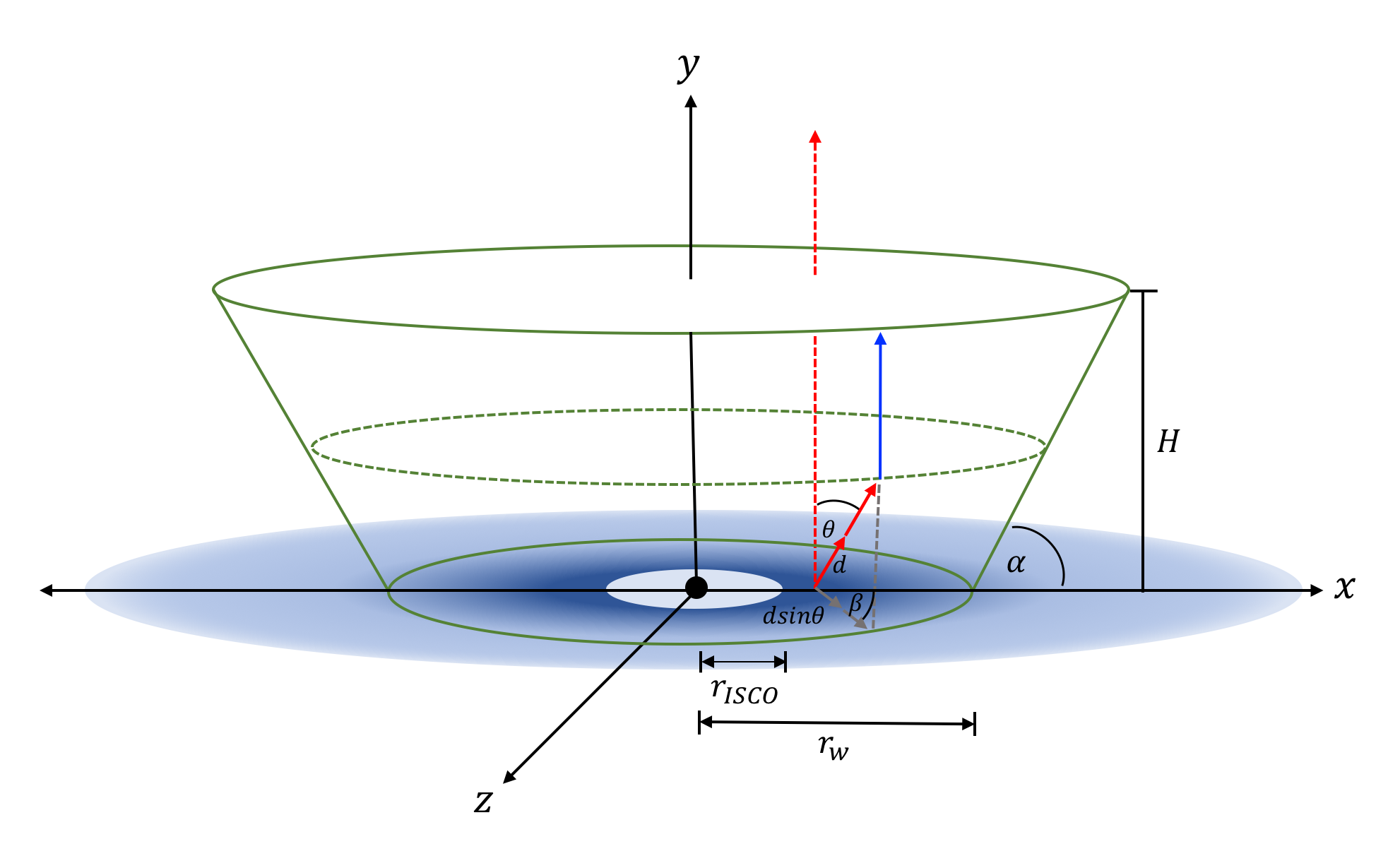}
    \caption{Schematic representation of the single scattering model. The blue shade indicates the plane of the accretion disc around the central black hole (black dot) in which the outflowing wind is launched from $r_{\rm w}$. The green line sketched the geometry of the wind set by the wind parameters $r_{\rm w}$, $H$ and $\alpha$  so that allow the opening funnel to be formed at the central region. The red dashed line indicate the hard photons that travel directly to the observer while the red solid line is the hard photons that travel for the distance $d$ and scatter at the wind surface; the latter becomes the soft photons (blue solid line) and arrive the observer later, making the photon time delay. The travelling direction of these photons could be defined by the parameters $\theta$ and $\beta$ (see text for detail).}
    \label{fig:singlesc}
\end{figure*}

For the single scattering model, the super-Eddington accretion around a Kerr black hole of mass $M_{\rm BH}$ is assumed. Thus the matter could orbit circularly down to the innermost stable circular orbit ($r_{\rm ISCO}$). Since most BHBs and SMBHs have been reported that they are likely to have very high spin ($a$ $\ga$0.9; \citealt{reynolds2021}), here we assume that the central BH has a maximum rotating rate so that the $r_{\rm ISCO}$ could be $\sim$1.23$r_{\rm g}$\footnote{$r_{\rm g} = GM/c^2$ where $G, M$, and $c$ are the gravitational constant, the BH mass, and the speed of light, respectively.} \citep{Thorne1974,Laor1991}. The model is also set to allow the outflowing wind with having the height $H$ and the angle $\alpha$ to be launched from the wind launching radius $r_{\rm w}$, so that it forms the opening funnel on the disc surface. The model geometry and parameters are shown in Fig.~\ref{fig:singlesc}.

Given the proposed model structure, the hard photons would be emitted anywhere from the annulus area with the inner radius and outer radius of $r_{\rm ISCO}$ and $r_{\rm w}$, respectively. We define the scattered photons as ones that emerge in the direction defined by the angle $\theta$ and $\beta$ and travel with the distance $d$ until hit and down-scatter at the edge of the outflowing wind becoming the soft photons, as shown in Fig.~\ref{fig:singlesc}. In addition, we simply define the hard, un-scattered photons as the photons that have $\theta$ = 0$^{\circ}$ so that they travel directly to an external observer without scattering with the wind. Therefore, if $\Delta d$ is the path difference between these two types of photons, their time delays ($\tau$) could be calculated from 
\begin{equation}
t = \Delta d/c.
\end{equation}
However, we are interested in the lags between two energy bands which are soft 0.3--1 keV and hard 1--7 keV bands. The response function, $\psi(t)$, determining the amount of scattered photons from the wind observed as
a function of time is produced. The Fourier form of the response, $\Psi(f)$, then is calculated and the phase difference between two energy bands is \citep{Cackett2014}
\begin{equation}
\phi(f) = \tan^{-1}\bigg{(}\frac{{\rm Im}(\Psi)}{1 + {\rm Re}(\Psi)}\bigg{)},
\end{equation}   
where ${\rm Im}(\Psi)$ and ${\rm Re}(\Psi)$ denote the imaginary and real parts of $\Psi(f)$. The lag-frequency spectrum of these two bands then are
\begin{equation}
\tau(f) = \frac{\phi(f)}{2\pi f}.
\end{equation} 
Negative and positive signs of the lag, in our definition, means the soft band lagging the
hard band, and the hard bend lagging the soft band, respectively.

Here, we also take into account the dilution effect in which the scattered photons could be diluted by, e.g., the direct emission from the outflowing wind in form of the soft photons having the same energy band with the down-scattered photons. Due to the dilution effects, the measured lags is always smaller than the intrinsic time lags and could lead to the error in measurement of the true source height \citep[see, e.g.,][and discussion therein]{Wilkins2013, uttley2014, Chainakun2015, Chainakun2016}. The dilution effect usually is parametrized by the refection fraction, $R$, defined as the ratio of (reflection flux)/(continuum flux) in the band of interest. However, to avoid the degeneracy of the model, we simply assume $R=1$ so the scattered photon flux to the direct emission photons flux in the soft band is equal while no contribution from the scattered photon flux in the hard band. Therefore, the free parameters in this model are limited to be $r_{\rm w}$, $\alpha$, $H$ and $M_{\rm BH}$. 

Using the geometry and conditions explained, the response function obtained from the model at different parameter values could be simulated and used to explain the observed time-lag data of the ULX. The photons from the inner disc is relatively hard in the way that it mainly contributes in the 1--7 keV band. Some of the disc photons are down-scattered from the disc wind producting the reflection flux in the 0.3--1 keV band. Hence, the soft energy band lags behind the hard band, resulting in the negative time lags in the lag-frequency spectrum. 

On the other hand, positive lags are also observed in both AGN and X-ray binaries, which are usually distributed to the mechanisms associating with the inward-propagating fluctuations along the disc ocuuring on the viscous timescales \citep[e.g.,][]{Kotov2001, Arevalo2006}. Here, we select to simulate the positive lags using a power-law model, as in, for example, \cite{Emmanoulopoulos2014} and \cite{Epitropakis2016}.

\subsection{Multiple scattering} \label{sec:Multiple scattering}

For comparison, time lags due to X-ray reverberation under multiple-scattering framework are also estimated. In this framework, we still assume the same accretion geometry with the previous model; however, this model allows the hard photons to scatter with the wind surface more than one time. We employ a simple lamp-post model in which an isotropic point source is on the rotational axis of the black hole, instead of allowing photon emission from the disc. In this way, the analytic solution of time taken by each photon that went along multiple-scattering within the wind funnel can be derived. 

\begin{figure}
	\includegraphics[width=\columnwidth]{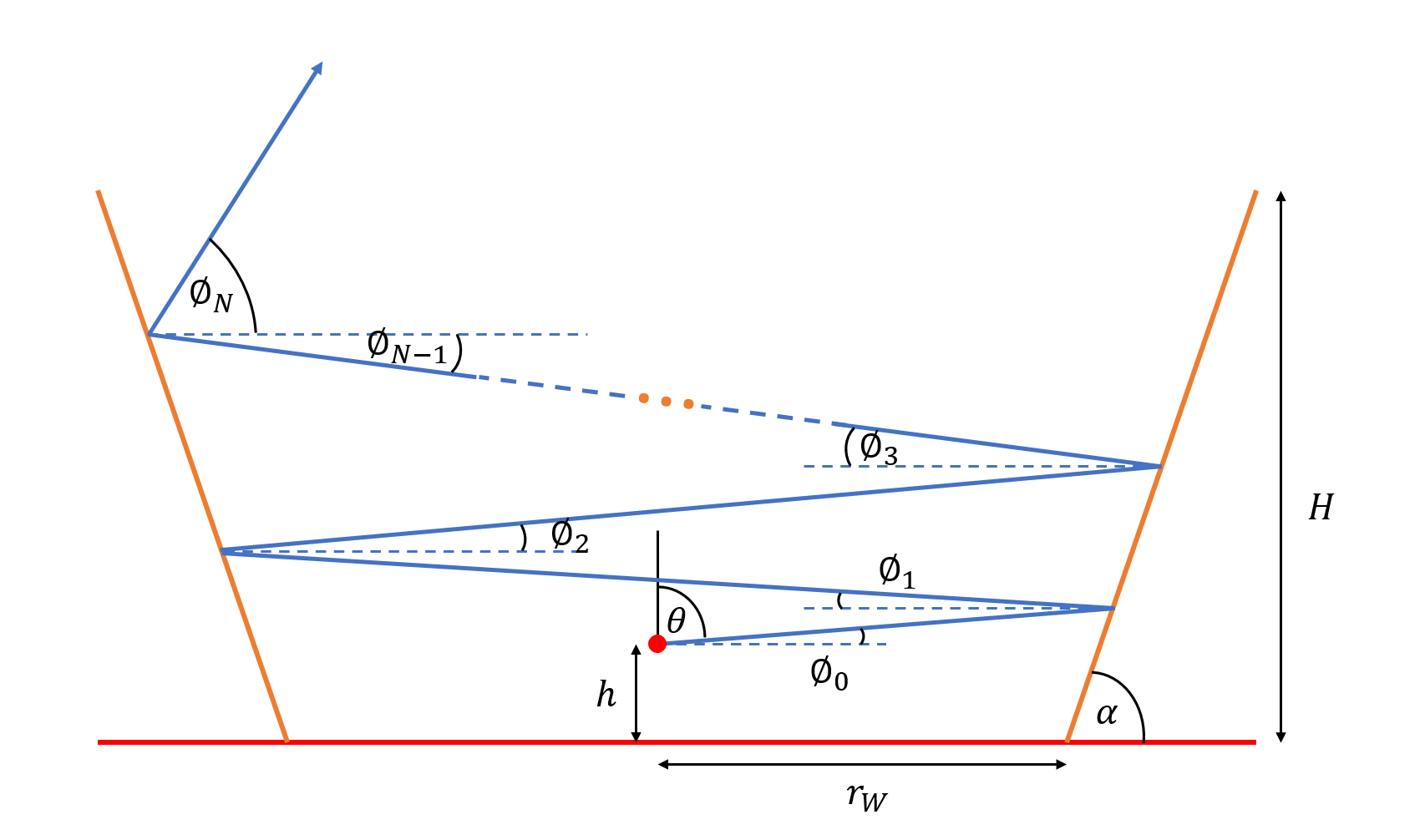}
    \caption{Schematic representation of the multiple scattering model. As for Fig.~\ref{fig:singlesc}, the wind (orange line) is launched from the disc (red line) starting from the point $r_{\rm w}$ as its geometry is defined by the parameters $r_{\rm w}$, $H$ and $\alpha$. The red dot indicates the position that photons are emitted. While a portion of the hard photons move to the observer directly, some photons are lagged as they could scatter with the wind surface multiple times before escape the wind (blue solid line); the travelling direction of the lagged photons could defined by the parameters explained in Section~\ref{sec:Multiple scattering}.}
    \label{fig-multsc}
\end{figure}

In this scenario, the source height is always fixed at $h=2r_{\rm g}$ that could be a representative of the X-ray emission from the innermost regions of accretion flow. The model parameters are the wind height ($H$), the wind launching radius ($r_{\rm w}$), and the angle between the accretion disc and the wind ($\alpha$), as same as the single scattering case. The geometry setup and essential parameters that are used to calculate time lags are presented in Fig.~\ref{fig-multsc}. 

Note that, for simplicity, we consider the multiple scattering of photons in two-dimensional flat space and omit the Doppler and relativistic effects so that an analytical solution of the light-travel time could be derived. The aim is to show a comparison of fitting results if two different scenarios are assumed: single and multiple scattering ones. Nevertheless, we remark that the light-travel time of a photon between each scattering should be relatively larger than the time it takes in travelling near black hole where relativistic effects are strong. Therefore, the flat space assumption (e.g., light travel in straight lines) could still be a good approximation to calculate the time delays due to multiple scattering of photons at the surface of large-scale outflowing wind. In fact, this assumption is fairy good approximation and usually employed when the size scale of direct emission is relatively small and the scatter or reflection takes place far from the central black hole \citep[e.g.,][]{Mahmoud2019, Chainakun2021}.

The photon scatter occurs in the way that the angle of incidence always equals the angle of reflection, measured with reference to the line perpendicular to the wind surface. The incident and reflection angle of each scatter is convert to the angle $\phi$ measured compared to the horizontal line, as illustrated in Fig.~\ref{fig-multsc}, so that we can write 
\begin{equation}
   \phi_n = \phi_0 + 2n(\pi/2-\alpha),
	\label{eq:x}
\end{equation}
where $\phi_0 = \pi/2-\theta$ and $n=1, 2, 3, ..., N$ is the $n^{\rm th}$ number of scattering and $N$ is the total number of times of scattering. $\theta$ is the emission angle of the source photons and $\alpha$ is the wind launching angle. 

The distance that a photon travels from the X-ray source to the wind before their first scattering is given by 
\begin{equation}
d_0 = \frac{h + r_{\rm w}\tan{\alpha}}{\cos{\phi_0}\tan{\alpha} - \sin{\phi_0}}.
\end{equation}
The distance that the photon travels between $n^{\rm th}$ and $(n+1)^{\rm th}$ scattering, except for $n = N$, is given by
\begin{equation}
d_n = \frac{2h + 2r_{\rm w}\tan{\alpha} + 2\sum_{i=0}^{n-1}d_{i}\sin{\phi_i}}{\cos{\phi_n}\tan{\alpha} - \sin{\phi_n}}.
\end{equation}
The distance for the photon travelling out of the wind after the last, $N^{\rm th}$ scattering is 
\begin{equation}
d_N = \frac{H - (h + \sum_{i=0}^{N-1}d_{i}\sin{\phi_i})}{\sin{\phi_N}}.
\end{equation}
Consequently, the total distance the photon travels before escaping the wind is
\begin{equation}
d_{\rm tot} = d_0 + \sum_{i=1}^{N-1}d_{i} + d_N,
\end{equation}
hence the corresponding total time the photon takes before escaping the wind can be written as 
\begin{equation}
t_{\rm tot} = d_{\rm tot}/c.
\end{equation}

To produce the response function in the multiple scattering scheme, we firstly determine the total number of scatter, $N$, for each emission photon from the source. This could be done by placing constraint for reaching the final scattering to be $\phi_{N} \geq \alpha$ or when the total height of the photon has already travelled is larger than $H$. Once the total number of scattering of each photon is known, we can calculate $d_{\rm tot}$ and $t_{\rm tot}$. Then, for a given geometry, the corresponding response function can be produced by assuming isotropic emission and summing the total number of photons as a function of $t_{\rm tot}$ substracted with the average time the direct photons take in travelling to the observer. Note that, similar to the single scattering model, the model free parameters are $r_{\rm w}$, $\alpha$, $H$ and $M_{\rm BH}$, with the assumed reflection fraction $R=1$. The lag-frequency spectrum between 0.3--1 and 1--7 keV bands is computed and the positive lags associating with the disc propagating-fluctuations are modeled in the form of a simple power-law.

\subsection{Characteristic of the models} \label{sec:Characteristic of the models}

\begin{figure*}
	\includegraphics[width=30pc]{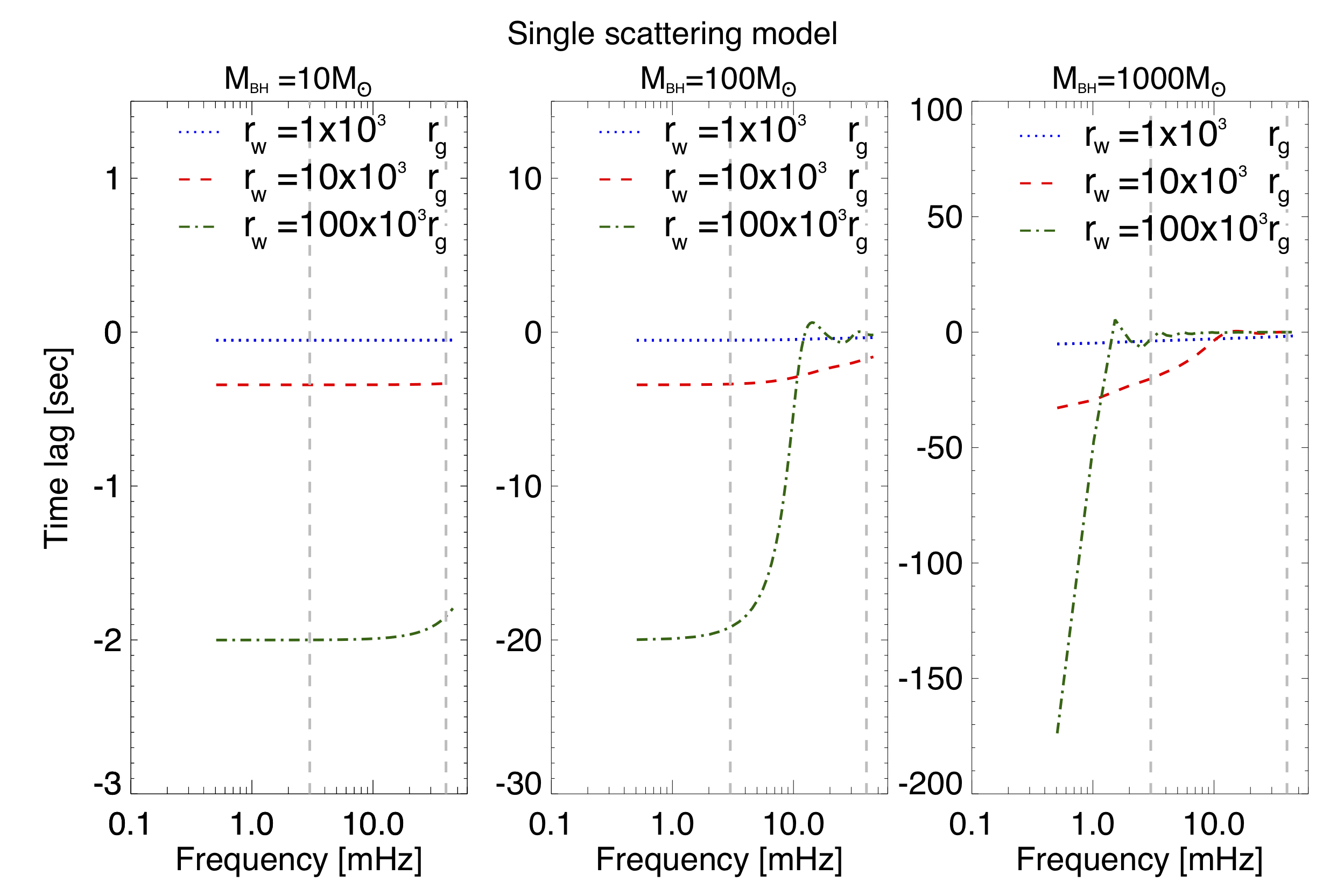}\\
	\includegraphics[width=30pc]{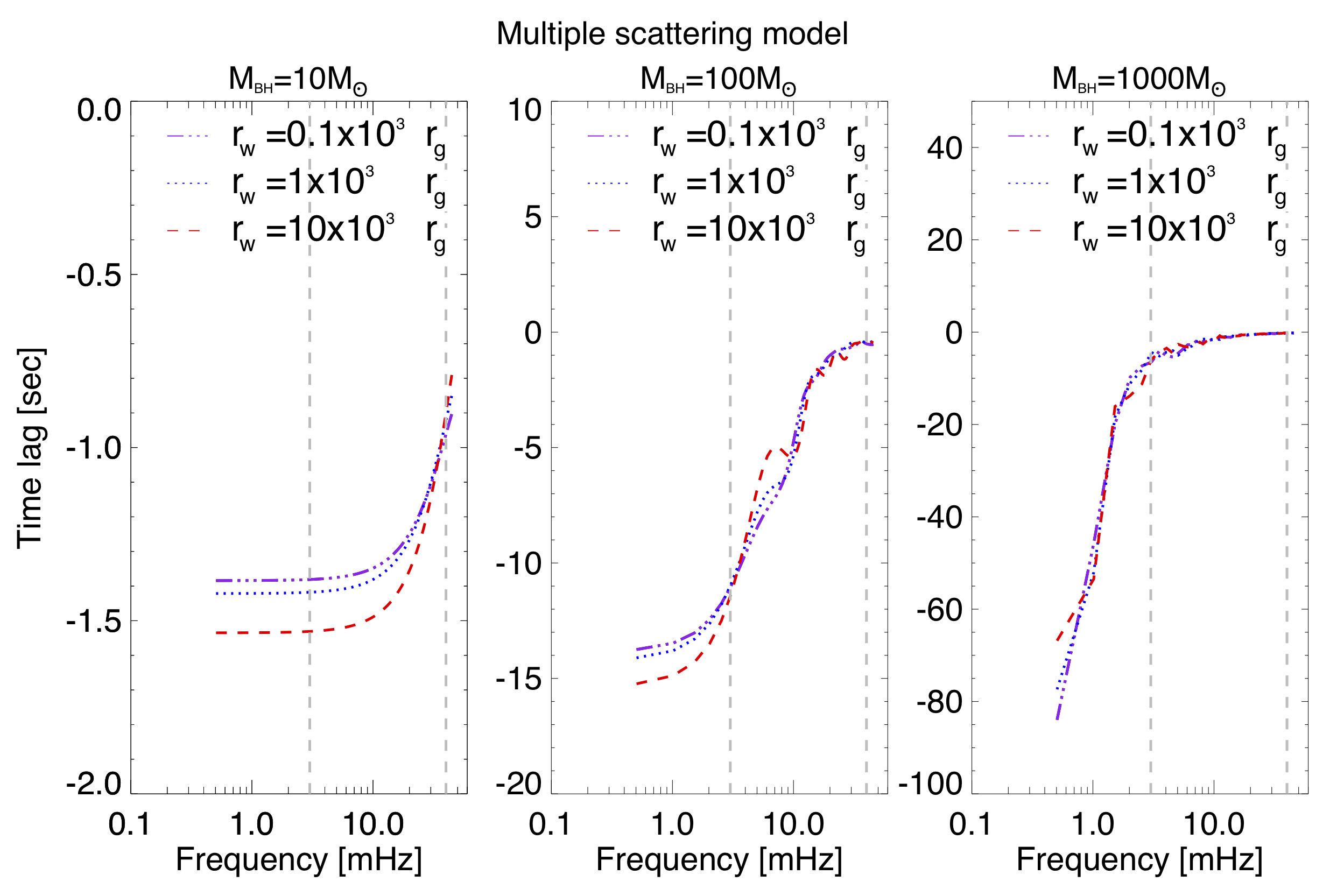}
    \caption{The time lag spectra varied as a function of the BH mass ($M_{\rm BH}$) and the wind launching radius ($r_{\rm w}$) for the single scattering model (top) and multiple scattering model (bottom); the wind height ($H$) and the wind angle ($\alpha$) are fixed at $10^{3}r_{\rm g}$ and 30$^\circ$, respectively. The vertical dashed lines indicate the observed frequency range of the data.}
    \label{fig:model_character1}
\end{figure*}

\begin{figure*}
	\includegraphics[width=30pc]{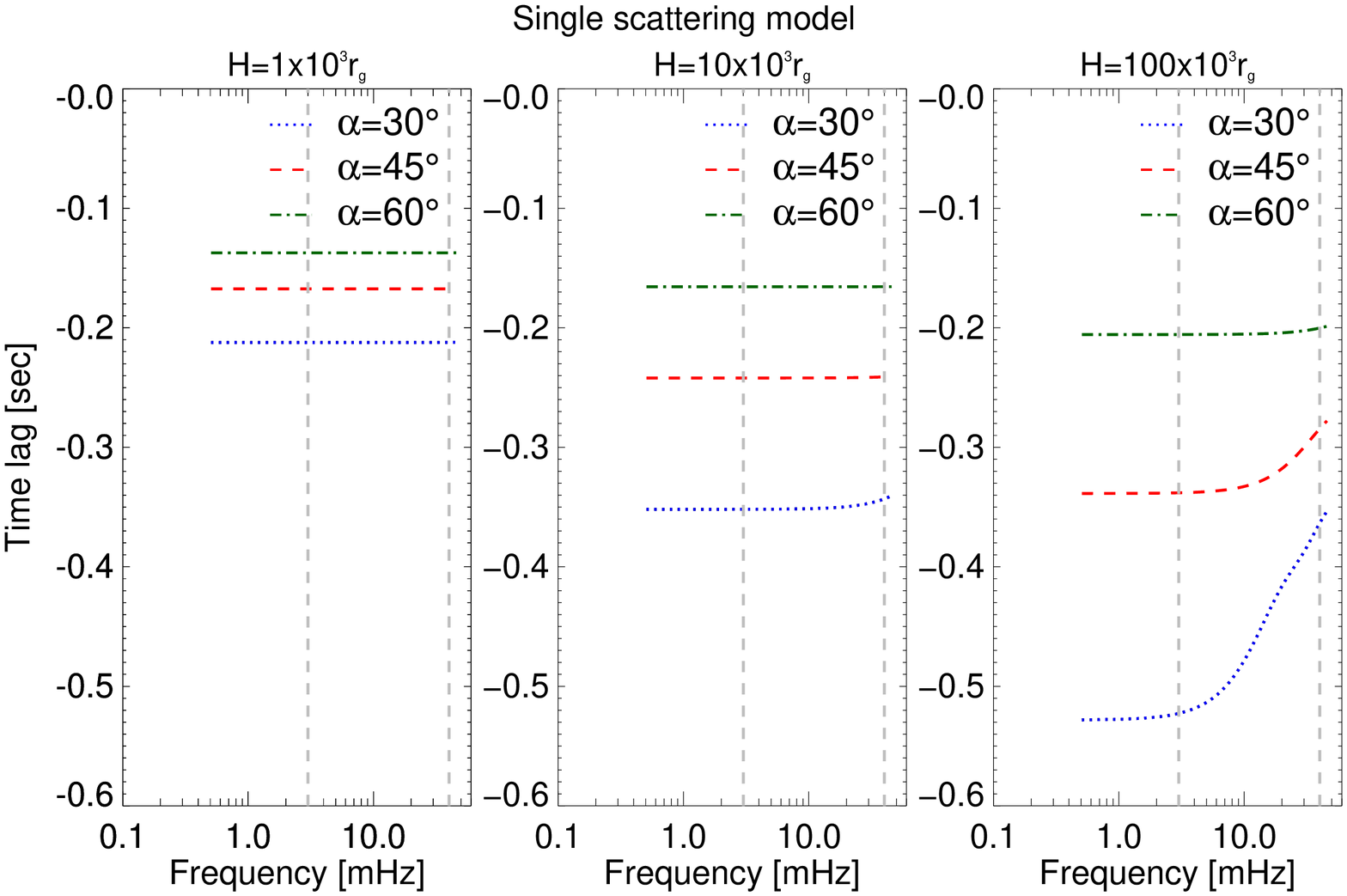}\\
	\includegraphics[width=30pc]{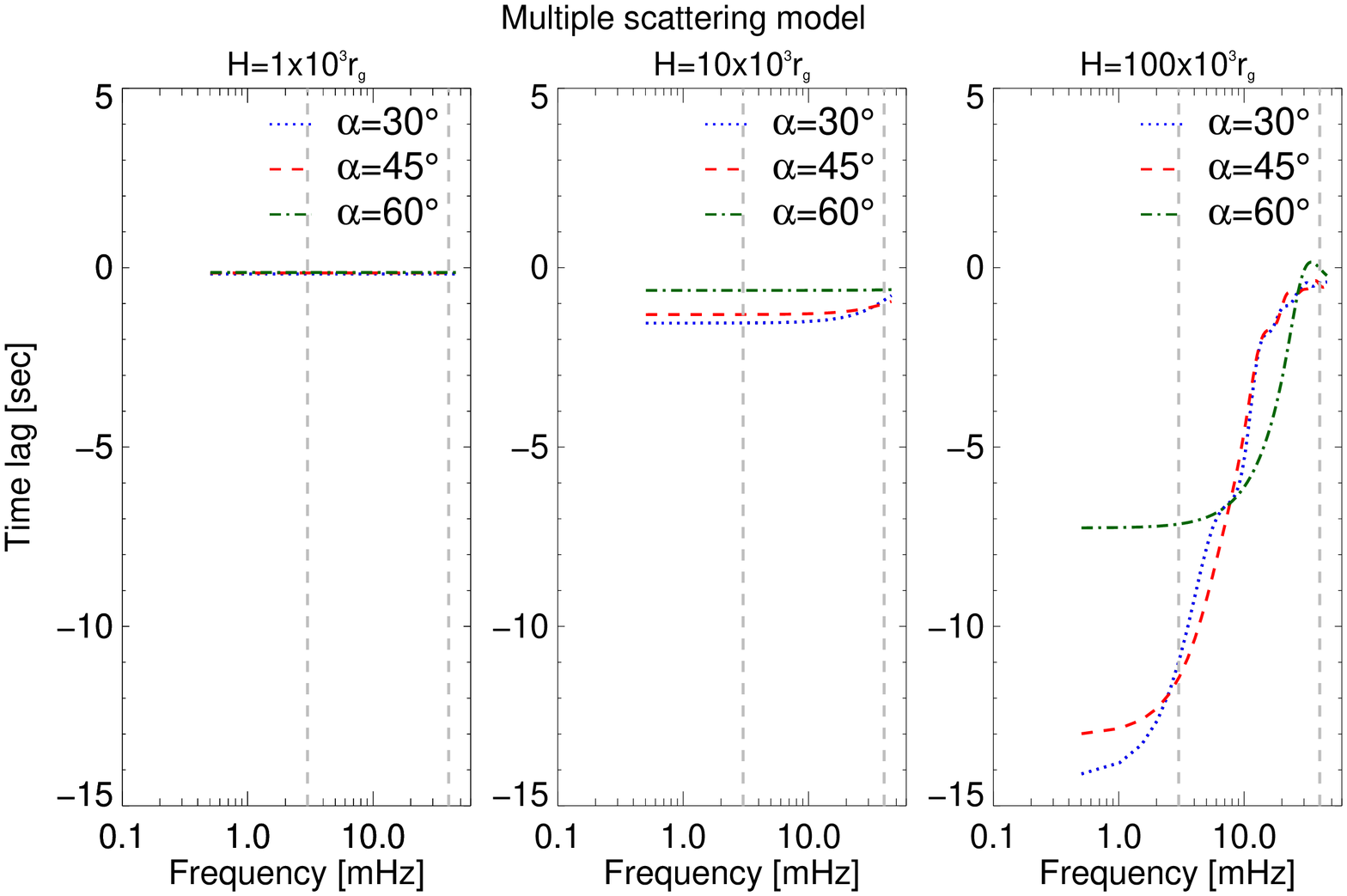}
    \caption{The time lag spectra varied as a function of the wind height ($H$) and the wind angle ($\alpha$) for the single scattering model (top) and multiple scattering model (bottom); the wind launching radius and BH mass are fixed at 1$\times$$10^{3}$$r_{\rm g}$ and 100 M$_{\rm \odot}$, respectively. The vertical dashed lines indicate the observed frequency range of the data.}
    \label{fig:model_character2}
\end{figure*}

To get the idea how the time-lag spectra change as a function of the model parameters,  the spectra at different values of $r_{\rm w}$, $\alpha$, $H$ and $M_{\rm BH}$ are plotted and illustrated in  Fig.~\ref{fig:model_character1} and Fig.~\ref{fig:model_character2}; here,  the range of frequency showed is between 0.5 - 50 mHz so that we can study them broadbandly while they also covers the observed frequency range of the data. Overall, both single scattering and multiple scattering models are consistent in which the lag spectra are scaled with the BH mass; as we might expect, the longer lag could be obtained if the BH is more massive (see Fig.~\ref{fig:model_character1}). In addition, the similar behaviour was also obtained from the increment of the other parameters: the wind launching radius $r_{\rm w}$ and the wind height $H$. Namely, the farther $r_{\rm w}$ and taller $H$ could result in longer time lag as the photons need to travel for longer distance until they escape the wind. However, given that the increment of the BH mass and the wind launching radius (as well as the wind height) could similarly produce the longer lag, it seems that there would be a degeneracy between these parameters; for example, the same lag amplitude could be obtained from either more massive BH with nearer wind, or smaller BH in which the wind is launched farther away. Fortunately, this is not the case if we consider the time-lag spectra: i.e., considering the lag in a range of frequency. In fact, although the same lag amplitude could be obtained, the spectral shape could be different, especially in the observed frequency range (see e.g. Fig.~\ref{fig:model_character1} top panel). Moreover, we also found that, as the BH mass increases, the lag spectra would be scaled in which the soft lag amplitude would begin to peak at lower frequency; for instance, the frequency is shifted from $\sim$10 mHz to $\sim$1 mHz as the BH mass increase from 10 M$_{\rm \odot}$ to 100 M$_{\rm \odot}$, in case of multiple scattering model (Fig.~\ref{fig:model_character1} bottom panel). These actually allow our models to be able to constrain the model parameters efficiently. In addition, for the spectral evolution as a function of the wind angle $\alpha$ (see Fig.~\ref{fig:model_character2}), it was found that, for both models, the longer lag could be obtained from lower wind angle. This is because for lower angle of the wind, the wind geometry would be flatter and the opening funnel is wider (i.e., the wind angle $\alpha$ is smaller) so that the photons would travel longer until hit the wind surface.

However, although the two models seem to share the  common characteristic, the models are also different, actually, due to the scattering mechanism defined for each model. Indeed, while the spectral evolution of the single scattering seems to be straightforward, in which increment of BH mass, $r_{\rm w}$, and $H$, and the decrement of $\alpha$ would result in higher lag amplitude, that of the multiple scattering model is quite complicated. In fact, we found that the lag amplitude of the multiple scattering model is more sensitive to the wind height (Fig.~\ref{fig:model_character2} bottom panel), whilst it seems to be less sensitive (or insensitive) to the distance from BH to the wind launching radius (Fig.~\ref{fig:model_character1} bottom panel). This could be because the wind height would be the major parameter, playing an important role to regulate the number of photon scattering; more number of scattering times could be obtained from taller wind. Moreover, the situation would be more complicated if we consider the interplay of the parameter $H$,
$r_{\rm w}$ and $\alpha$. For example, at fixed values of $H$ and $\alpha$, the increment of $r_{\rm w}$ would result in the longer time lag as the photons would travel longer and scatter multiple times; however, as the parameter $r_{\rm w}$ continues to increase, the time lag could be reverse which the shorter time lag would be obtained since the number of scattering is reduced; the wind is not sufficiently high to produce more number of scattering for most photons. Similarly, for the fixed values of $H$ and $r_{\rm w}$, the decrement in the parameter $\alpha$ would seems to boost the time lag at the beginning as the photons would travel longer; later this would act to reduce the number of scattering since high fraction of photons that  scatter first time with the wind surface would then escape the wind so that the time lag amplitude is reduced. 

\begin{figure*}
	\includegraphics[width=\columnwidth]{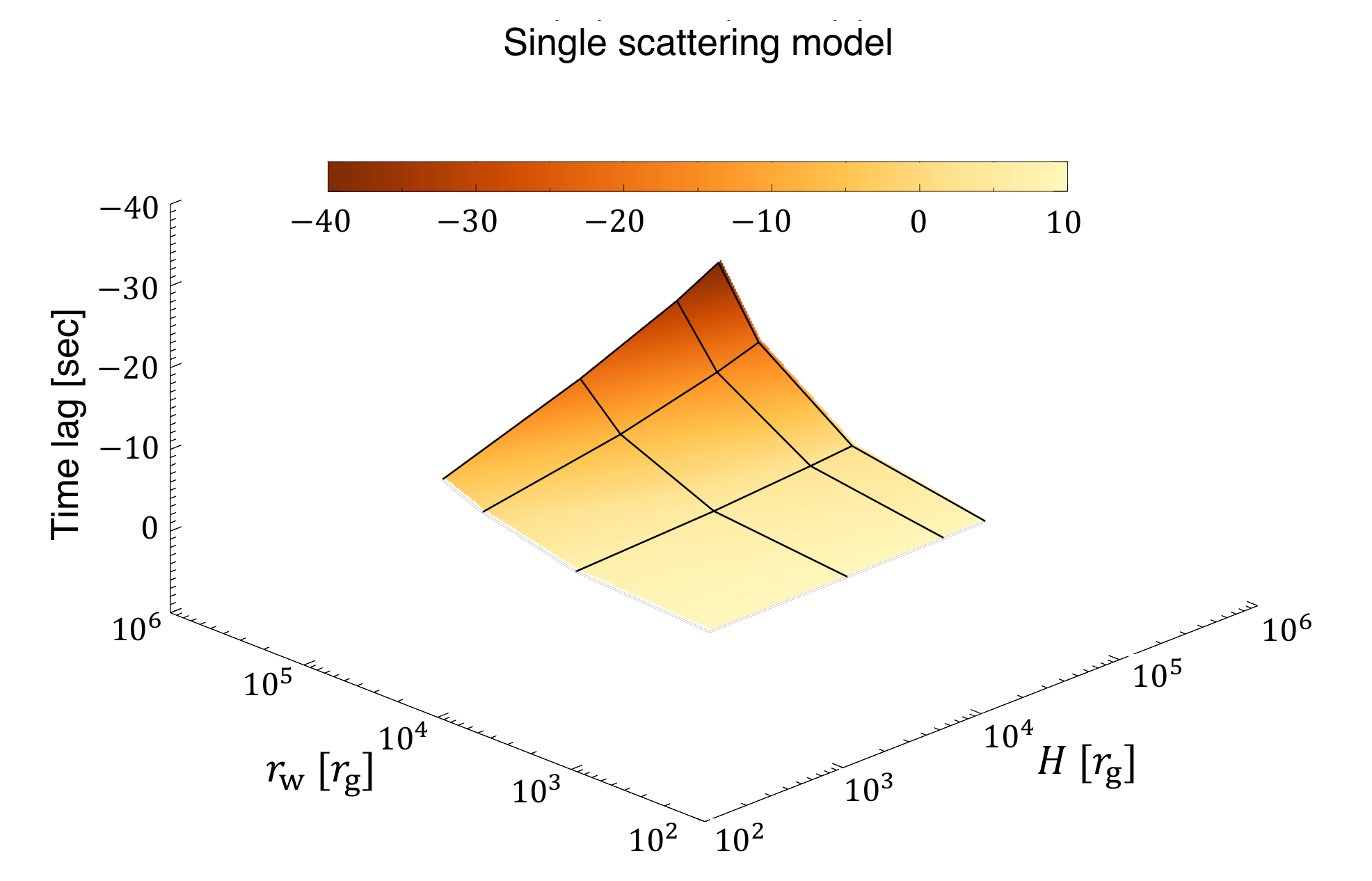}~\includegraphics[width=\columnwidth]{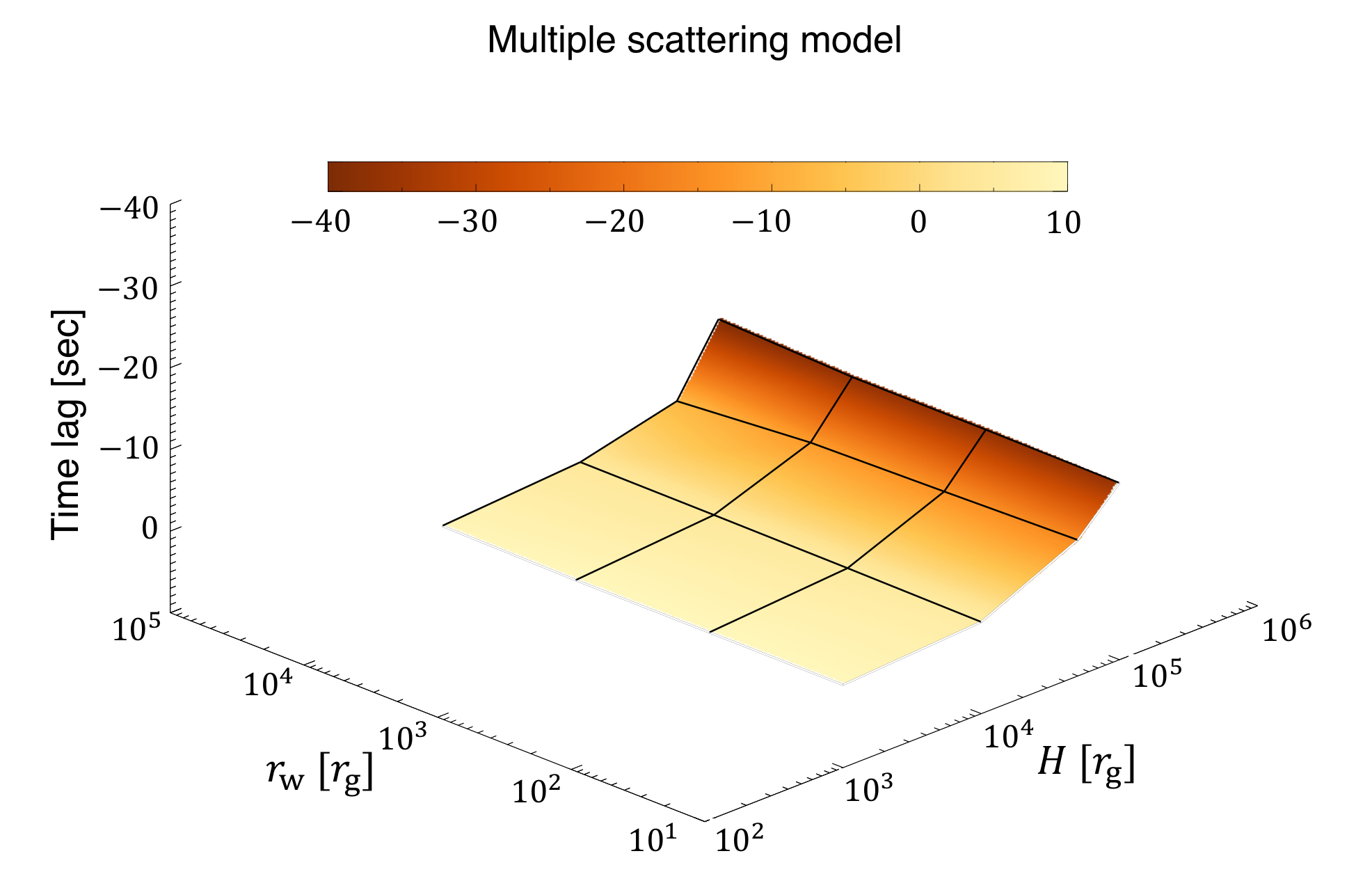}
    \caption{The surface plots illustrate how the soft lag amplitude at 3 mHz varies as a function of the wind launching radius ($r_{\rm w}$) and the wind height ($H$) for single scattering model (left) and multiple scattering model (right); the BH mass and the wind angle are fixed at 100 M$_{\odot}$ and 30$^{\circ}$, respectively.}
    \label{fig:model_character3}
\end{figure*}

Finally, we also created the surface plot to roughly estimate the effect of $r_{\rm w}$ and $H$ to the amplitude of the soft lag as shown in Fig.~\ref{fig:model_character3}. Obviously, for the BH mass in the order of 100 M$_{\odot}$, the values of $\ga$ 10$^{4}$$r_{\rm g}$ for both $r_{\rm w}$ and $H$ is required to obtain the lag of a few seconds (i.e. about the same level of the lag amplitude observed in our data), in particular for the single scattering model. However, for the multiple scattering model,  as mentioned above, it seems the lag is more sensitive to the wind height, so the values of $H \ga$10$^{4}$$r_{\rm g}$ is required to explain the observed lag spectra.

\section{Results} \label{sec:Results}

We fitted the lag spectra of the ULX NGC 5408 X-1 by the models explained in Section~\ref{sec:models} using the capability of \textsc{xspec} version 12.10.0 \citep{arnaud1996}. To exclude low signal-to-noise data bins from the analysis, we selected to fit the lag spectral data over the frequency range of 3 -- 40 mHz. The FTOOL script \textsc{ftflx2xsp}\footnote{\url{https://heasarc.gsfc.nasa.gov/lheasoft/ftools/headas/ftflx2xsp.html}} was used to convert the time-lag spectra file into standard PHA files and their corresponding RSP files which are readable by \textsc{xspec}. The models were imported into \textsc{xspec} by adding user-defined models, i.e. the table models,\footnote{\url{https://heasarc.gsfc.nasa.gov/docs/heasarc/caldb/docs/memos/ogip\_92\_009/ogip\_92\_009.pdf}} in which the model parameters are the free parameters described in Section~\ref{sec:models}.  Furthermore, since the mass of central black hole should not change significantly on the time scale of years that this ULX was observed, the analysis of low, medium and high luminosity bin spectra were performed simultaneously in which the parameter $M_{\rm BH}$ of each lag spectrum was tied together in order to obtain only single best fitted value. All best fitted parameters were calculated using $\chi^{2}$ minimisation method and their uncertainties are quoted with error bars that is equivalent to the 90 per cent confidence interval. In addition, since we fitted the luminosity bin spectra simultaneously, to be able to examine if the models could explain some spectrum better than the others, here, we also report the {\it individual $\chi^{2}$} of each time-lag spectrum; basically, this is the $\chi^{2}$ value obtained from each spectrum which contributes to the final, best fitted $\chi^{2}$ of the fitting; in fact, the summation of all individual $\chi^{2}$ values result in the final, best fitted $\chi^{2}$ value.

The analysis began by fitting the spectra with the pure single scattering model; the result is shown in Table~\ref{tab:sg_fitting}. It is quite clear that the model could not well reproduce the data (reduced $\chi^{2}$ $\sim$ 2)\footnote{Here, we defined the reduced $\chi^{2}$ as the best fitted $\chi^{2}$ over degree of freedom.} for both results from the averaged and luminosity bin spectral fittings, excepting for the high luminosity spectrum in which the individual $\chi^{2}$ is $\sim$ 5, significantly lower than the other luminosity bin spectra. In addition, it was also found that some model parameters could not be constrained in which the upper and/or lower limits, especially the parameter $H$ could not be calculated. Moreover, the BH mass obtained from averaged spectral bin and luminosity-based spectral bins are completely inconsistent, i.e. 1000~M$_\odot$ vs 90~M$_\odot$. These point that the model might be less practical and still need some modification.


\begin{table*}
\begin{center}
\end{center}
\caption{\label{tab:sg_fitting}Fitting results obtained from the pure single scattering model.}
\centering
\begin{threeparttable}
\begin{tabular}{@{}*{7}{c}}
\hline
Spectral bin 	&$ r_{\rm w} [\times 10^3r_{\rm g}]$ 				&$\alpha [^\circ]$		&${H} [\times 10^3r_{\rm g}]$  				&~~$M_{\rm BH} [\rm M_\odot]$ 					& Individual $\chi^{2~a}$ & Best fitted $\chi^2 /$ d.o.f.$^{~b}$\\
\hline
~~~Averaged & $44.73\substack{+3.22\\-3.06}$	 		&$30.00\substack{+8.53\\~}$				& $1000.00^*$								&$958.24\substack{+15.65\\-3.36}$	& -& $13.52/6$\\
~~~~~Low   		& $1.00\substack{+0.72\\-1.00} $	 							&$56.40\substack{+33.60\\-16.06}$				& $1000.00^*$								&							&$10.44$ &\\
~~~Medium		& $4000\substack{~\\-737.77}$							&$30.00^*$				&$166.98\substack{+481.26\\-166.98}$								&$89.83\substack{+4.58\\-4.78}	$	&$23.36$ & $38.96/20$\\
~~~~~High		&$992.76\substack{+482.68\\-238.32}$		&$35.78^*$				&$499.21\substack{~\\-231.00}$								&							&$5.16$ &\\
\hline

\end{tabular} 
         \begin{tablenotes}
         \item \textit{Note.} $^{a}$The individual $\chi^{2}$ value of each time-lag spectrum contributing to the best fitted $\chi^{2}$ values (see text for details) -- no value reported for the averaged spectrum since it was fitted alone. $^{b}$The best fitted $\chi^{2}$ statistic, and the number of degree of freedom obtained from the fitting. $^{*}$The parameter uncertainty could not be constrained. For some best fitting values that either upper or lower limit cannot be constrained, the error bar was reported only in one direction.
         \end{tablenotes}
\end{threeparttable}
\end{table*}

\begin{figure*}
	\includegraphics[width=\columnwidth]{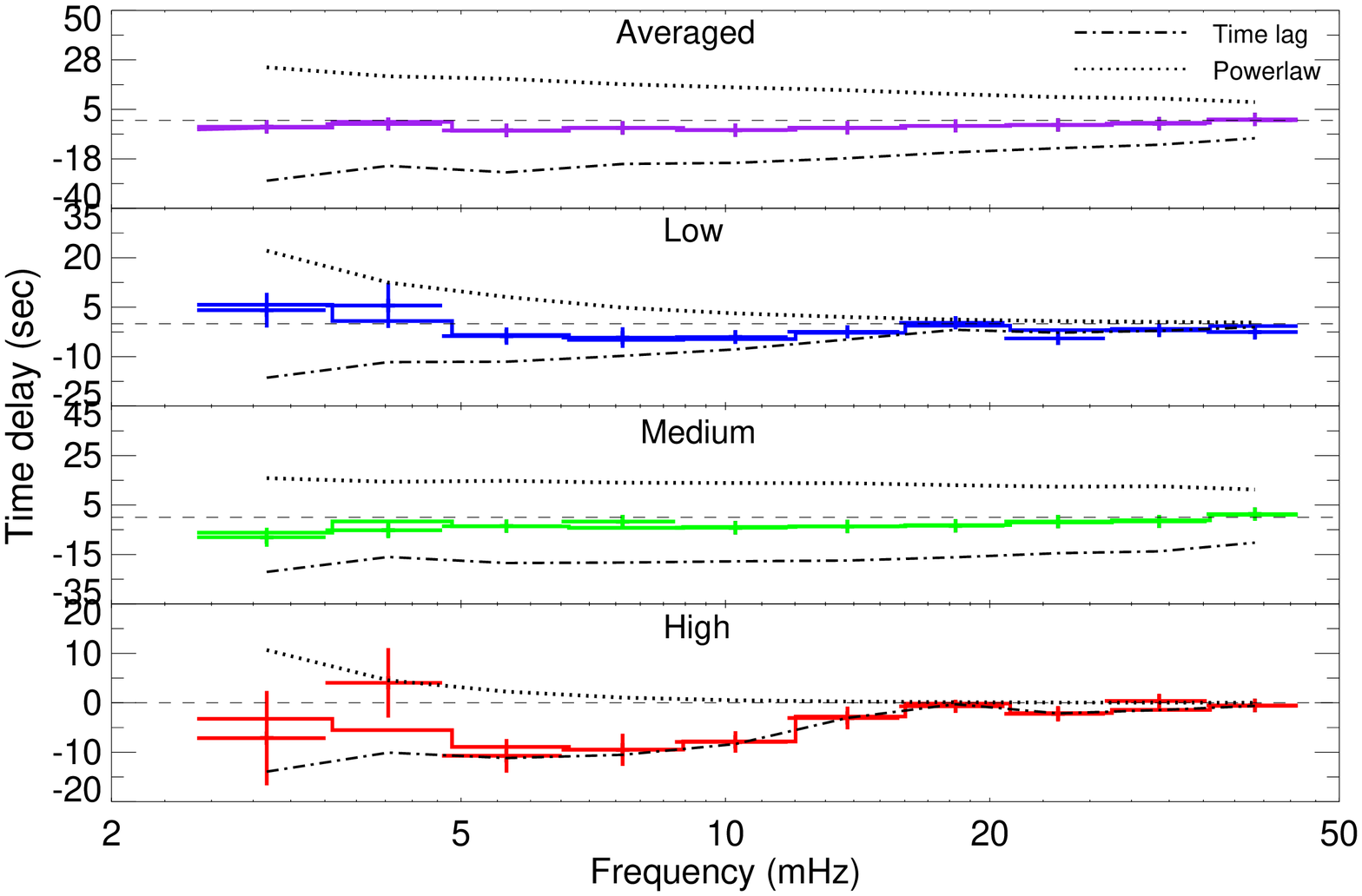}~\includegraphics[width=\columnwidth]{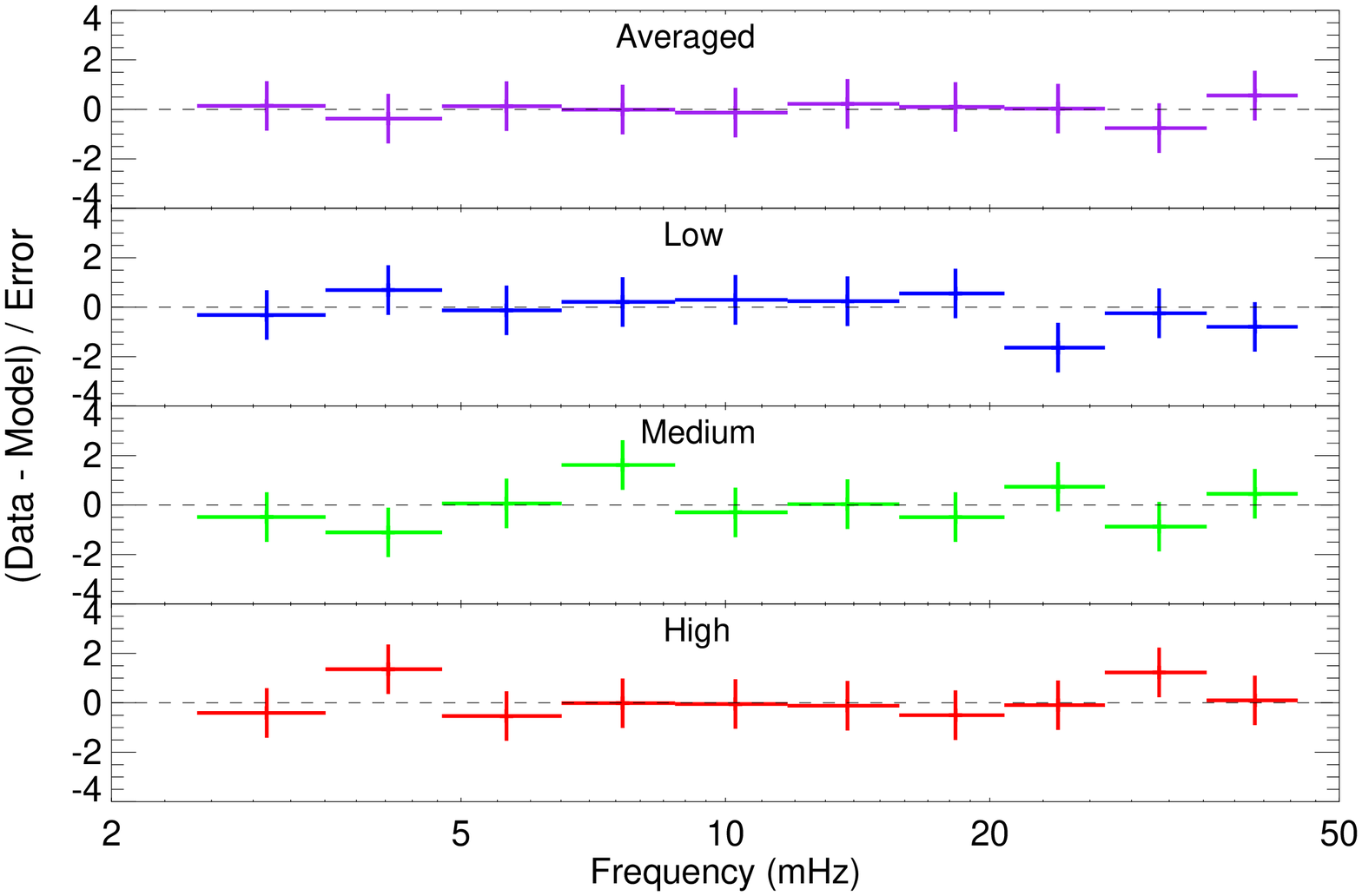}
    \caption{The time-lag spectra best fitted by the single scattering + power-law model (left) and their corresponding fitting residue (right).}
    \label{fig:sg+po_fitting}
\end{figure*}

Thus, we modified the single scattering model by adding the power-law component. Indeed, this is often used to account roughly for the hard lag actually found in BHBs as well as AGN (e.g. \citealt{miyamoto1988,nowak2000,walton2013}) commonly interpreted as the inward propagation along the disc, frequently named as propagating fluctuation \citep{lyubarskii1997}. The motivation for adding this is also hinted by the low frequency hard lag existing in the low luminosity bin spectrum. Indeed, this mechanism is also expected in the accretion flow of super-Eddington accretion disc such as that of ULXs \citep{middleton2015}.  The result from the modified model is shown in Table~\ref{tab:sg+po_fitting} top panel and Fig.~\ref{fig:sg+po_fitting}. By adding the power-law component, it is clear that the model improved the fitting results significantly for both averaged and luminosity bin spectra; for the averaged spectrum, the best fitted $\chi^{2}$ is 1.14 while for the luminosity bin spectra, the individual $\chi^{2}$ is $\sim$ 4 -- 6 for each spectrum. Note that for the high luminosity bin spectrum, the value of individual $\chi^{2}$ did not improve significantly, comparing to the pure scattering model. This might imply that the lag data in this bin do not require a contribution (or require just a small contribution) from the power-law component to explain them, supporting by the substantially low value of the power-law normalisation parameter ($\la$ 1 $\times$ 10$^{-3}$).  Indeed, considering best fitted values of the normalisation parameter, it seems that the lag data would require significantly the power-law component to describe only for the low and medium luminosity spectra (as well as for the averaged spectrum). In addition, the values of power-law slope seems to range between 0 -- 2. Thus, these could suggest that the lag spectra would need different level of the power-law contribution to describe the hard lag and/or the dilution effect.

\begin{table*}
\caption{\label{tab:sg+po_fitting}Fitting results obtained from the single scattering + power-law model.}
\centering
\begin{threeparttable}
\begin{tabular}{@{}*{10}{c}}
\hline
\multirow{2}{*}{Spectral bin} &  \multicolumn{3}{c}{Single scattering} && \multicolumn{2}{c}{Power-law}		&\multirow{2}{*}{~~$M_{\rm BH} [\rm M_\odot]$ }		&   Individual$^{a}$		&   Best fitted$^{b}$
\\\cmidrule{2-4}\cmidrule{6-7}

&$ r_{\rm w} [\times 10^3r_{\rm g}]$ 		&$\alpha [^\circ]$				&  ${H} [\times 10^3r_{\rm g}]$  	 	 && $\Gamma$ 					&Norm. 							& 		&$\chi^{2}$ 	&$\chi^2 /$ d.o.f. \\
\hline

\multicolumn{10}{c}{\it{Free all varied parameters}} \\
Averaged &$500.74\substack{+18.90\\-3.09}$		&$45.30\substack{+45.30\\~}$	&$9.99\substack{+18.09\\-3.09}$		&&$0.28\substack{+0.22\\-1.28}$		&$3.72\substack{+8.19\\-2.90}$		&$108.33\substack{+256.41\\-21.30}$	&	-&$1.14/4$\\
~~~~~Low   	& $1332.91\substack{~\\-411.79}$	&$30.00\substack{+27.16\\~}$	& $276.29\substack{+631.23\\-228.70}$	&&$1.48\substack{+0.46\\-0.63}$		&$3.07\substack{+73.83\\-3.04}\times10^{-3}$ 			&				&$4.46$\\
~~~Medium	& $450.36\substack{+86.89\\-28.51}$	&$30.00\substack{+60.00\\~}$	&$6.38\substack{+25.16\\-1.20}$		&&$0.00\substack{+0.32\\-1.00}$		&$12.37\substack{+0.54\\-4.11}$	&$84.00\substack{+3.40\\-7.69}$		&$5.93$		&$14.49/14$\\
~~~~~High	&$1198.22\substack{+433.82\\-360.90}$	&$56.02\substack{+21.57\\~}$	&$480.95\substack{~\\-195.51}$	&&$2.34\substack{+0.58\\-3.34}$		&$9.76\substack{+293.22\\-9.76}\times10^{-6}$ 	&			&$4.11$\\

\hline

\multicolumn{10}{c}{\it{Freeze $\alpha$ at 45$^\circ$}} \\
Averaged & $500.74\substack{+73.31\\-32.99}$			&   45.00 &$9.99\substack{+15.58\\-3.20}$			&&$0.28\substack{+0.13\\-1.28}$		&$3.75\substack{+11.39\\-1.99}$				&$99.67\substack{+272.04\\-13.29}$	&	-&$1.13/5$\\
~~~~~Low   	& $1453.23\substack{~\\-531.06}$	 & 45.00 & $239.13\substack{~\\-199.99}$	&&$1.25\substack{+0.51\\-0.85}$		&$12.91\substack{+86.66\\-12.90}\times10^{-3}$ 			&	&$5.54$\\
~~~Medium	& $459.74\substack{+105.56\\-48.26}$		&  45.00   &$13.83\substack{+23.90\\-6.48}$			&&$0.14\substack{+0.21\\-1.14}$		&$6.94\substack{+4.73\\-4.76}$				&$85.42\substack{+7.47\\-9.90}$		&$7.59$		&$18.16/17$\\
~~~~~High	&$1312.60\substack{+388.72\\-468.07}$	& 45.00      &$765.13\substack{~\\-412.80}$		&&$3.09\substack{+0.78\\-4.09}$		&${1.24\substack{+406.55\\-1.24}\times10^{-7}}$ 						&								&$ 5.03$\\

\hline
\end{tabular}
         \begin{tablenotes}
         \item \textit{Note.} $^{a}$The individual $\chi^{2}$ value of each time-lag spectrum contributing to the best fitted $\chi^{2}$ values (see text for details) -- no value reported for the averaged spectrum since it was fitted alone. $^{b}$The best fitted $\chi^{2}$ statistic, and the number of degree of freedom obtained from the fitting. For some best fitting values that either upper or lower limit cannot be constrained, the error bar was reported only in one direction.
         \end{tablenotes}
\end{threeparttable}
\end{table*}


Given the goodness of fits obtained, the best fitted parameters could reveal us the properties of the accreting system. One of the advantages is that the model could constrain the mass of BH powering the accretion. From the averaged spectrum, the model suggest the mass of $\sim$108 M$_{\rm \odot}$ while the luminosity bin spectra suggests the 84 M$_{\rm \odot}$ BH. Interestingly, the values of BH mass obtained from the averaged and luminosity bin spectra are well consistent within the error bars which they could lie in the regime of MSBH (or low mass IMBH). However, the value of mass obtained from the luminosity bin spectra is much better constrained than that obtained from  the averaged spectrum, i.e. the error bars of the former are much smaller than that of the latter. This could suggest that grouping the time lag data into the luminosity bins could provide more accurate value of the BH mass, probably because the lag spectra are different in their properties, and so the averaged spectra might not represent well the time-lag properties of the ULX.

However, for the parameters indicating the geometry of the wind, i.e. $H$, $\alpha$ and $r_{\rm w}$, it seems that the model could not constrain well these values in most cases, probably because each individual time-lag spectral quality -- i.e. the signal-to-noise -- is low. Nevertheless, considering the best fitted values, for the averaged spectrum, the magnitude of wind height and wind launching radius are $\sim$ 10$^{4}$ -- 10$^{6}r_{\rm g}$. This could suggest that the hard photons would scatter with the wind at the distance much further away from the central BH. In addition, for the model geometry implied by the best fitted parameters from the luminosity bin spectra, we found the similar result with that of the averaged spectra, i.e. being consistent within the statistical uncertainty; however, it is obvious that the parameter ranges obtained from the luminosity bin spectra, especially that of the low and high luminosity bins, are much larger, comparing to the averaged spectrum. Furthermore, the parameters  $H$ and $r_{\rm w}$ seem to vary unnaturally in which the values drop when the luminosity changes from low to medium luminosity bin but then swap to increase when the luminosity rises to high luminosity bin. These might suggest that the model would require the higher quality of the time-lag spectra to constrain these information.

To better constraint the wind geometry parameters in the luminosity bin spectra, we attempted to reduce the number of free parameters in the model. We began by fixing the wind angle ($\alpha$) at the median value, i.e. 45$^\circ$. This is also supported by the best fitting value of this parameter obtained from averaged time-lag spectrum. The fitting result is shown in the bottom panel of Table~\ref{tab:sg+po_fitting}. Overall, by decreasing 1 free parameter for each spectrum, the individual $\chi^{2}$ was increased by $\sim$1 for low, medium and high luminosity bins; we regard that this is an acceptable fitting result. In fact, it is interesting that the mass of BH obtained from the model still did not changed (i.e. $\sim$85 M$_{\rm \odot}$). However, considering the remaining free parameters related to the wind geometry, i.e. $H$ and $r_{\rm w}$, the best fitted values are similar to the previous model while no significant improvement in the error bars were obtained. In addition, we also tried to fix the wind angle at other values, i.e., 30$^\circ$ and 60$^\circ$; the significant improvement in the best fitted statistic or the better constraint on the varied parameters was not found. These might imply that the wind angle might not be an important parameter that explain the time-lag spectra of the ULX. Indeed, the parameter might significantly effect the time-lag amplitude at lower frequency, out of the observed frequency range (see Fig.~\ref{fig:model_character2}). We also further analysed to see how the parameter $H$ affect the fitting result by trying to fix the value of $H$ at 50$\times10^{3}r_{\rm g}$ and 500$\times10^{3}r_{\rm g}$ during fitting the data. It was found that the best fitting $\chi^{2}$ increased by $\sim$15 and $\sim$19, respectively, suggesting that the parameter $H$ significantly affects the time lag and would be required to change with the increasing luminosity.

\begin{table*}
\caption{\label{tab:ms+po_fitting}Fitting results obtained from the multiple scattering + power-law model.}
\centering
\begin{threeparttable}
\begin{tabular}{@{}*{10}{c}}
\hline

\multirow{2}{*}{Spectral bin} &  \multicolumn{3}{c}{Multiple scattering} && \multicolumn{2}{c}{Power-law}		&\multirow{2}{*}{~~$M_{\rm BH} [\rm M_\odot]$ }		&Individual$^{a}$		&Best fitted$^{b}$
\\\cmidrule{2-4}\cmidrule{6-7}

&$ r_{\rm w} [\times 10^{3}r_{\rm g}]$ 		&$\alpha [^\circ]$				&  ${H} [\times 10^{3}r_{\rm g}]$  	 	 && $\Gamma$ 					&Norm. 							& 		&$\chi^{2}$ 	&$\chi^2 /$ d.o.f. \\
\hline

Averaged		&$10.00\substack{\\-10.00}$	&$7.97\substack{+5.89\\-7.97}$						&$99.95\substack{~\\-55.68}$		&&$0.12\substack{+0.05\\~}$	&$7.22\substack{+6.20\\-1.68}$	&$142.11\substack{+401.54\\-48.89}$	&	&$1.33/4$\\
~~~~~Low   	& $0.020\substack{+0.0219\\-0.009}$&$27.36\substack{+10.51\\-6.53}$	&$94.22\substack{~\\-14.53}$	&&$0.31\substack{+0.14\\-0.20}$		&$3.32\substack{+2.63\\-1.42}$ 					&				&$2.75$\\
~~~Medium	& $0.322\substack{+0.144\\-0.138}$	&$16.10\substack{+2.95\\-5.55}$	&$100.00\substack{~\\-11.75}$	&&$0.00\substack{+0.09\\-1.00}$		&$11.91\substack{+0.54\\-3.47}$	&$83.45\substack{+5.76\\-6.43}$		&$8.32$		&$14.36/14$\\
~~~~~High	&$0.015\substack{+0.032\\-0.015}$	&$30.31\substack{+14.91\\-7.63}$	&$87.39\substack{~\\-10.41}$	&&$0.03\substack{+0.17\\-1.03}$		&$10.49\substack{+2.31\\-5.76}$ 						&			&$3.29$\\
\hline
\end{tabular}
         \begin{tablenotes}
         \item \textit{Note.} $^{a}$The individual $\chi^{2}$ value of each time-lag spectrum contributing to the best fitted $\chi^{2}$ values (see text for details) -- no value reported for the averaged spectrum since it was fitted alone. $^{b}$The best fitted $\chi^{2}$ statistic, and the number of degree of freedom obtained from the fitting. For some best fitting values that either upper or lower limit cannot be constrained, the error bar was reported only in one direction.
         \end{tablenotes}
\end{threeparttable}
\end{table*}
\begin{figure*}
	\includegraphics[width=\columnwidth]{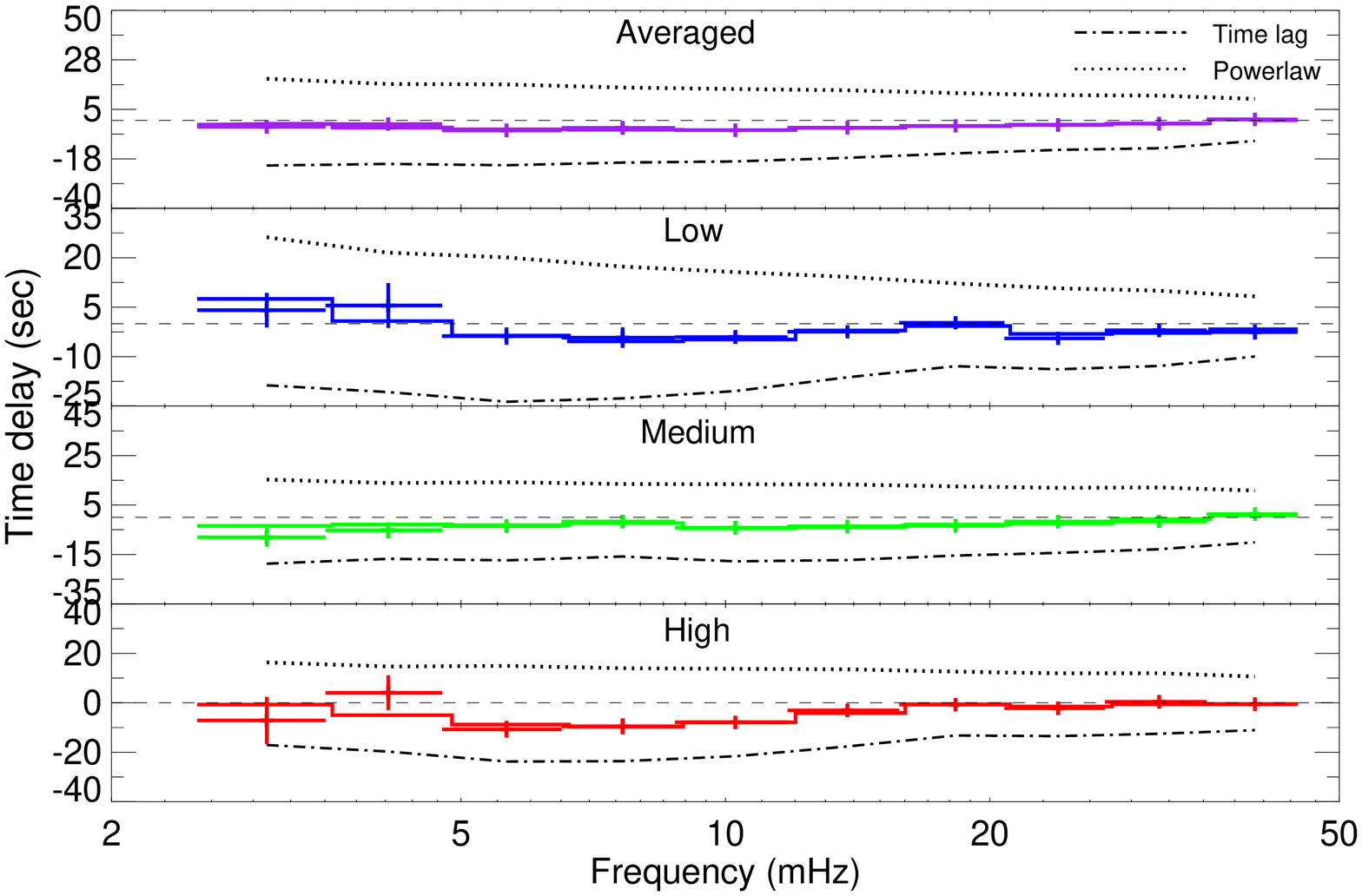}~\includegraphics[width=\columnwidth]{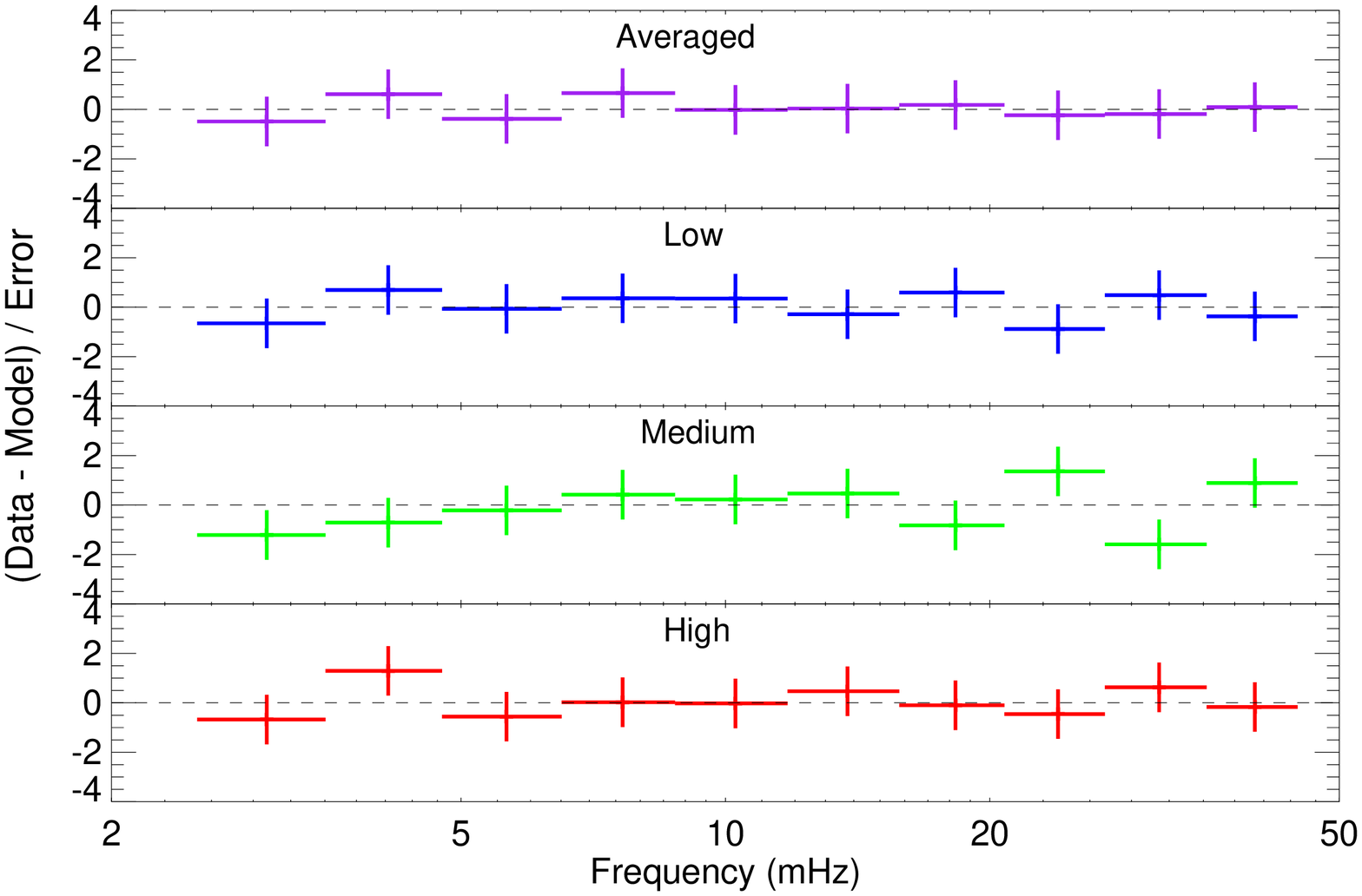}
    \caption{The time-lag spectra best fitted by the multiple scattering + power-law model (left) and their corresponding fitting residue (right).}
    \label{fig:ms+po_fitting}
\end{figure*}


Moreover, we alternatively attempted to model the time-lag data using the multi-scattering model proposed in Section~\ref{sec:Multiple scattering}. Similar to the single scattering model, we also added the power-law component to account for the hard lag, probably due to the propagation fluctuation. Thus, the number of model free parameters are the same with that of the single scattering model: i.e. d.o.f. = 4 and 14 for averaged spectrum and luminosity bin spectra, respectively. The fitting result is shown in Table~\ref{tab:ms+po_fitting} and Fig.~\ref{fig:ms+po_fitting}.  It was found that the model fitted the data well and yields roughly the same best fitting statistic to that of the single scattering model: reduced $\chi^{2}$ $\sim$ 0.33 and 1 for the averaged spectrum and luminosity bin spectra, respectively. However, we note that for the medium luminosity bin spectrum, the model worse describe the spectrum comparing to the low and high luminosity bin ($\sim$60 per cent of total $\chi^{2}$ were contributed by this spectrum) because the model could not duplicate well the spectrum, in particular the strong hard lag feature at lower frequency bins.

In addition, we found that the multi-scattering model seems to constrain better the varied parameters -- especially for the parameter $\alpha$, $H$ and $r_{\rm w}$ implying the geometry of the wind-- comparing to the single scattering model. For the averaged spectrum, the $H$ and $r_{\rm w}$ are in the order of 10$^{4}$ - 10$^{5}r_{\rm g}$, which is comparable to that of the single scattering model. However, the lags produced by the multiple scattering model seem to be more sensitive to the wind angle, $\alpha$, than those produced by the single scattering model, especially for larger $H$ (see also Fig.~\ref{fig:model_character2}). This might be because the significant change in the amplitude of the lags likely depends on the number of scattering that is controlled by the wind geometry. In the averaged spectrum case, the wind angle was found to be $\sim$8$^\circ$, lower than that of the single scattering scenario. This might suggest that the spectrum required the model with low number of scattering to explain (see Section~\ref{sec:The disc geometry and its evolution} for more detail). However, for the luminosity bin spectra, we found that the wind geometry obtained from the low and high luminosity bin spectra are consistent while we skip to interpret the result from the medium luminosity spectrum due to the high value of fitting statistic explained above. Thus, this model seems to not suggest the evolution of wind geometry with luminosity.

Furthermore, considering the parameter $M_{\rm BH}$ obtained from the best fitting model, we found that the model suggests the BH mass of $\sim$140 M$_{\rm \odot}$  and $\sim$80 M$_{\rm \odot}$ for the averaged spectrum and luminosity bin spectra, respectively; indeed, the values are consistent within the error bars. However, similar to that of the single scattering model, the BH mass obtained from the luminosity bin spectra is better constrained than that of the averaged spectrum. Moreover, we also found that mass of BH suggested by the multiple scattering model is consistent with that suggested by the single scattering model. This might imply that the data required that unique value of BH mass, regardless of the models used and how the spectra are binned.

Finally, we also found that the power-law component are also required to explain the time-lag data in the multiple scattering model, indicated by the non-zero values of normalisation  parameter. However, the similar power-law contribution is required to explain the averaged spectrum and luminosity bin spectra; the best fitting power-law indexes and normalisations in all spectral cases are not significantly different. Thus, it seems that there is no evolution of the power-law component has been suggested by the multi scattering model.

\section{Discussion} \label{sec:discussion}

The reverberation of the ULX NGC 5408 X-1 were discovery and studied previously by several works:  e.g. \cite{Heil2010, demaco2013,garcia2015}. In this work, we revisited the reverberation of the ULX NGC 5408 X-1 using the data from archival observations of {\it XMM-Newton} observatory. Two types of time-lag spectra has been created: i) time averaged spectrum and ii) luminosity bin spectra; the purpose of the latter are to study the evolution of the time lag as the luminosity changes. All created spectra have been fitted by the single scattering model and the multiple scattering model which they are similarly assumed that the hard photons would scatter at the edge of the outflowing wind launched from the accretion disc; the models yielded an acceptable fitting results. 

The modelled lags here in the single scattering case are comparable to those reported in previous literature. For example, the lag amplitude is approximately in the same order of what estimated by \cite{demaco2013} assuming the comparative size of the wind geometry. The trend of the modelled lags is also in agreement with, e.g., \cite{Wilkins2013, Cackett2014, Emmanoulopoulos2014, Epitropakis2016} in the way that increasing the distance from the X-ray source to the reflector (regardless of the accretion disc or the wind) increases the amplitude of the lags. Our lags in all cases also scale with the black hole mass, as expected \citep{DeMarco2013new}. However, to the best of our knowledge, exact calculations and treatments for the multiple scattering scenario have not been investigated in detail before in previous studies. In the following sub-sections, we will interpret and discuss important keys obtained from Section~\ref{sec:Results}.

\subsection{The variability of the time-lag spectra} \label{sec:The variability of the time-lag spectra}

It has been reported in previous studies (e.g. \citealt{demaco2013,garcia2015}) that the ULX NGC 5408 X-1 showed the  soft lag of $\sim$5 - 15 seconds peaking at the frequency of $\sim$5 mHz, and it is likely to switch to hard lag at lower frequency.\footnote{The similar behavior has also been detected in the ULX NGC 4559 X7 \citep{pintore2021}} Although, in this work we limited the lowest frequency of the analysis down to 3 mHz to gain high signal-to-noise data, we still found the similar  time-lag properties in the ULX spectra. For the averaged time-lag spectrum, it exhibits the soft lag of $\sim$ -5 seconds at lower end of the spectrum, suggesting that, averagely, the ULX have a soft reverberation lag as a main feature over our analysed frequency range. However, considering the luminosity bin spectra, the evolution of the ULX reverberation could be revealed (see e.g. Fig.~\ref{fig:sg+po_fitting}). At low luminosity, the ULX pronounces soft lag down to $\sim$5 mHz frequency and then switch to the hard lag at lower frequency. As the luminosity increases, there is no sign of hard lag at the low frequency end and the entire low frequency reverberation lag is soft, as suggested by the medium and high luminosity bin spectra. This might explain why there is the low statistical significant, hard lag detected at low frequency, reported by \citet{garcia2015}, since the averaged time-lag spectra would average over and dilute the soft and hard lag features from different luminosity. Here, using the luminosity to bin the data, we show that the low frequency lag might evolve from hard lag to soft lag as the luminosity increases. We, however, caution that this argument is limited to the availability of observational data and specific energy bands of interest that are used to calculate time lags. More observations with longer exposure time will be really needed to probe the lags to lower frequencies, in order to robustly confirm the absence of the hard lags at low frequencies in high luminosity bin spectra. The physical interpretation of the spectral evolution will be discussed in Section~\ref{sec:The disc geometry and its evolution}

Furthermore, regardless of the physical mechanism resulting in the variability of low frequency lag, binning and combining the time lag data using the luminosity could obviously result in the better constraint of the BH mass parameter in the models: i.e. the error bars is smaller comparing to the case of averaged spectrum (see Section~\ref{sec:Results}). This might imply that the variability of the time lag spectra might be one key to constrain the mass of compact object powering the ULX.

\subsection{The mass of BH powering the ULX}

The BH mass is one of critical parameters which was able to be estimated in this work. Regardless of the reverberation models  -- single versus multiple scatterings -- or spectral binning methods -- averaged versus luminosity bins -- were applied to the analysis, the results point out the similar scenario in which the BH mass with 90 per cent confidence interval might be $\sim$80 - 500 M$_{\rm \odot}$, ranging from the regime of high mass MsBHs to that of low mass IMBHs. Interestingly, the results from our models are well consistent with the ULX mass obtained from the dynamical constraint using VLT observations, in which the ULX mass of $\la$ 510 M$_{\rm \odot}$ has been reported (\citealt{cseh2013}; see also \citealt{cseh2011}). However, our result is still ambiguous in which the ULX might harbour  MsBH or IMBH. In fact, this is topic that is still active for debating at the moment. While  several timing property studies -- including the narrow band feature (i.e. QPO) -- are likely to imply that the ULX NGC 5408 X-1 might be powered by IMBH with the mass  of $\sim$1000 M$_{\rm \odot}$\citep{strohmayer2007,strohmayer2009a,pasham2012}, these values of mass seems to be more massive, comparing to the estimation from our models, at least by a factor of two. 

On the other hand, the ULX mass infers directly from the luminosity suggest the MsBH scenario (e.g. \citealt{strohmayer2007,gladstone2009}), consistent with our result. In fact, the observed timing properties, including the QPO, of the ULX could be alternative explained in context of the super-Eddington accretion onto a MsBH, similar to that has been found in the BHB GRS 1915+105 and some Narrow-Line Seyfert 1s \citep{middleton2011}. Furthermore, the soft lag feature detected in the ULX could also be explained in the framework of super-critical accretion onto a stellar mass object \citep{middleton2015}, in which primary, hard photons might scatter or propagation through the outflowing wind, becoming the lagged, soft photons \citep{garcia2015}. Thus, our result seems to favor the scenario of super-Eddington accretion onto MsBH. In fact, if we considering only the fitting results from that of the luminosity bin spectra, it was found that the BH mass is better constrained in which the mass with 90 per cent confidence interval could be $\sim$75 -- 90 M$_{\rm \odot}$, i.e. falling in the regime of MsBH. With the constrained BH mass in the order of $\sim 10 $--$ 100$ M$_{\odot}$, the observed negative lags in the frequency range of interest described by the model are the true soft lags, not just the phase wrapping produced by mathematical artefacts (see also Fig.~\ref{fig:model_character1}--\ref{fig:model_character2}). If we presume that the time-lag spectral variability of the luminosity bin spectra is genuine, and intrinsic to the ULX (we will discuss the possibility and reliability of this point in the next sub-section), this could strongly suggest that the ULX is likely to be powered by a MsBH, not a IMBH. If true, importantly, our reverberation models could be another effective, indirect method to estimate the ULX mass with exhibiting the reverberation feature.

Finally, we note that while several ULXs has been confirmed to be powered by pulsating NS \citep{bachetti2014,furst2016,israel2017a,israel2017b,carpano2018,sathyaprakash2019,rodriguez2020}, there has not been an evidence of pulsating signal detected from the ULX NGC 5408 X-1. However, given the our modelling results, it seems that the scattering models with the NS mass scenario could not explain well the time-lag spectra, probably constrained by the spectral shape and features (see Section~\ref{sec:Characteristic of the models}).  Indeed, given that the metallicity of the ULX host galaxy NGC 5408 is relatively low: $\sim1/10$ Z$_{\rm \odot}$ \citep{mendes2006,grise2012}, it is likely to form the BHB system of MsBHs hosting ULX in this galaxy (e.g. \citealt{mirabel2017}). Therefore, merging our result with the low metallicity galactic environment, it might be suggested that the ULX BH mass could be likely in the regime of MsBH.

\subsection{The disc wind geometry and its evolution} \label{sec:The disc geometry and its evolution}

In this work, two reverberation models were applied to study the disc wind geometry and its evolution of the ULX.  While the two models assumed the similar geometry, they are different by the times that the hard photons scatter with the wind surface, before escaping the wind. Given that both models could explain the time-lag spectra equally well, suggested by the best fitting $\chi^{2}$ values (see Section~\ref{sec:Results}),  the question here could be how many times that the photons scatter, i.e. which model could explain better the lag spectra for this ULX? The answer to this might be implied by the parameter $\alpha$ of the multiple scattering model. Indeed, the result from the average spectrum suggest the $\alpha$ value, i.e. the wind angle, of $\la$14$^\circ$, comparing to that of the single scattering model which the value is $\sim$45$^\circ$. In addition, since the model simply assumed that the photon reflected angle is the same with the incident angle. This could suggest that the spectra requires low number of scattering, probably one time or few times, in the multiple scattering model. Thus, this seems that both models suggest the similar scenario in which the number of photon scatters would be very low. 

The spectroscopic data of the ULX NGC 1313 X-1 indicated that the wind might be located at $\le$ 50$r_{\rm g}$ \citep{pinto2016}. However, referring to the model fitting results, it has been suggested that the hard photons would scatter with the wind at the distance of $\sim$ 10$^{4}$ -- 10$^{6}$ $r_{\rm g}$, further away from the central BH. In addition, if we assume the ULX BH mass of 80 M$_{\rm \odot}$ following the fitting result, given the ULX X-ray luminosity, the ULX might be accreting matter at $\sim1-2\frac{\dot{M}}{\dot{M}_{\rm Edd}}$; the spherisation radius could be at $\sim$$\frac{5}{3}$$\frac{\dot{M}}{\dot{M}_{\rm Edd}}$ $\sim$ 2-3$r_{\rm g}$ \citep{poutanen2007}. In fact, this value is $\sim$ 4-5 order of magnitude lower than our estimated location that the photons reflect off the wind. Furthermore, in \citet{kara2020}, it has been estimated that the wind located at spherisation radius of 100$r_{\rm g}$ would have very high electron density ($n_{\rm e}$ $\sim$ 10$^{20}$ cm$^{-3}$) so that the number of photon scattering at this region would be up to $\sim$ 10$^{4}$ times. This value is extremely high, especially comparing to the number of scattering obtained from our models. However, it has been still unclear that the reverberation lags observed in various ULXs would associate with the same mechanism. In case of the ULX NGC 5408 X-1, it could be interpreted that the observed reverberation could be mainly from the single scattering (or few scatterings) of hard photons with the extended wind that is much further away from the BH, $\sim$ 10$^{4}$ -- 10$^{6}r_{\rm g}$ \citep[e.g.][]{middleton2014,sutton2014,alston2021}. Also, this reverberation region is in agreement with that is expected by \citet{garcia2015}. We note that in this case, the physical meaning of the parameter $r_{\rm w}$ could be alternatively interpreted as the radius or distance from the central BH that the hard photons would reflect with the wind that is further away. 

In addition, the evolution of the disc wind geometry of the ULX could be interpreted from the variability of the model parameters best fitting to the luminosity bin spectra. If we consider the evolution of parameters from all spectra -- low, medium and high luminosity -- it was found that the evolution seems to be unreasonable. For example, the values of $r_{\rm w}$ from the single scattering model is highest at low luminosity, and lowest at medium luminosity; as the luminosity continues to increases to high luminosity, the $r_{\rm w}$ becomes high again. However, since the low luminosity spectrum is dominated by the hard lag, the reverberation would be much diluted. The parameter obtained from this spectrum might be highly uncertain and we will skip this spectrum from the interpretation. Therefore, if we consider only the medium and high luminosity bin spectra, it was found that, for the single scattering model, the wind parameters -- $r_{\rm w}$ and $H$ -- are likely to increase, broadly consistent with the theoretical prediction \citep[e.g.][]{poutanen2007}. However, for the multiple scattering model, the evolution, especially that of $r_{\rm w}$, seems to disagree with the theoretical prediction: i.e. decrease. This might be because the model is weighted by the outlier data of hard lag at the frequency of $\sim$4 mHz in the high luminosity bin spectrum,  so the magnitude of soft lag is underestimated in the model. Removing this data point would increase the value of soft lag in the model so that the parameter $r_{\rm w}$ would be increased and be consistent with theory. Obviously, the high quality data would be required to track the evolution of the disc wind in the future work.

We also note that, in principle, the reflection fraction, $R$, in which we froze to 1 could be changed across different disc-wind geometries. The larger $R$ is expected when the wind is large, steep, and close to the black hole (i.e. larger $H$, larger $\alpha$, and lower $r_{\rm w}$). The effects of $R$ is to dilute the lags while the frequency where the phase wrapping occurs does not change. In fact, by varying $R$ between 0.5--1.5, the amplitude of the lags changes only with a factor of $\lesssim 2$ \citep[e.g.][]{uttley2014}. The uncertainty on $R$ then may slightly affect the constrained $r_{\rm w}$ and $H$ here, but it should not alter the central mass scenario here that is likely to fall in the MsBH regime.   

Finally, it would be worth to mention the detection of 115 days periodic modulation in the ULX which might be interpreted as precession of the inner-disk/jet \citep{strohmayer2009b,foster2010}. If this is the case, the precession could result in the difference in angle of inclination respected to the observer's line of sight. In case that the ULX emission is anisotropic (i.e. beaming, especially, due to the wind opening funnel), our luminosity bin spectra might imply the difference in angle of inclination.  If so, the model parameter $\alpha$ should play an important role to improve the fitting. For the single scattering model result, we have showed that varying the parameter $\alpha$ does not help improve the fitting result significantly (see Section~\ref{sec:Results}). On the other hand, for the multiple scattering model result, although we could constrain the parameter $\alpha$ from the luminosity bin spectra, we found that there is no trend for the evolution from low to high luminosity. Additional better quality time-lag spectra would be required to model the spectra with the precession scenario.

\section{Conclusions} \label{sec:conclusion}

Among a few ULXs that exhibit the reverberation signal, in this work, we revisited the reverberation data of the ULX NGC 5408 X-1. The aim of the study is to estimate the accretion geometry and the BH mass that is powering the ULX. The time-lag spectra were created and binned using two different criteria: the time averaged spectral bin and luminosity-based spectral bin; the latter one was to track the spectral variability as a function of luminosity as well as accretion rate. Here, two ULX reverberation models -- single scattering and multiple scattering models -- were proposed to explain the time-lag properties of the observed time-lag spectra. Both models explained the spectra equally well and they consistently provide the key results as follows.

\begin{itemize}
    \item Regardless of the models, the result from averaged time-lag spectrum suggests that the BH mass powering the ULX could be $\sim$80 -- 500 M$_{\rm \odot}$, implying either MsBH or IMBH ULX. However, the result from the luminosity bin spectra point out that the central object could be MsBH of mass $\sim$75 -- 90 M$_{\rm \odot}$.
    \item The modelling results have suggested that the wind geometry is much extended in which the photons could be reflected and down scattered at the distance of $\sim$ 10$^{4}$ -- 10$^{6}$ $r_{\rm g}$, much further away from the central BH.
    \item Although the two models are different by the photon scattering times within the wind opening funnel, they, however, consistently suggest that the observed soft-lag photons could scatter with the outflowing wind only single time or a couple of times before escape the wind.
    \item The time-lag spectral variability of the ULX has been detected in which it changed from the hard to soft reverberation lags as the luminosity increased. Unfortunately, the data signal-to-noise is not sufficient to allow the models to constrain the evolution of the wind geometry with the spectral variability. However, the models seem to roughly suggest that the wind is more extended as the ULX luminosity increases.
\end{itemize}
 
Given above information, it has been found that reverberation mapping is one of helpful techniques to study ULX accretion geometry and mechanism as well as the mass of central compact object. Crucially, by grouping the time-lag data by the level of ULX luminosity, the central object mass could be constrained efficiently. Moreover, this should provide an opportunity to study the evolution of the outflowing wind, which is key to understand an accretion at super-critical rate. However, given the limited number of the ULX data available, further observations would be required to allow the models to probe this. Last but not least, the reverberation signatures could also be imprinted in the Power Spectral Density (PSD) profiles \citep[e.g.][]{Emmanoulopoulos2016, Papadakis2016, Chainakun2019new}. Recently, \cite{Chainakun2021new} applied the Machine Learning (ML) technique to extract the information of the X-ray source height from X-ray reverberation features that appeared in the PSD of AGN. The source geometry predicted by the ML model seems to be very accurate even in the case that the reflection fraction is very small, so does the amplitude of reverberation lags. Keeping in mind that the quality of data of many ULXs may not be sufficient to obtain meaningful constraints from the timing analysis \citep[e.g.][]{sutton2013}, it is interesting to see if the ML technique could overcome these issues by providing an accurate prediction of the geometry of the system from some of these data where the interpretable reverberation features may be very subtle. Applying the ML technique to study reverberation lags in ULXs in a systematic way is planned for the future.

\section*{Acknowledgements}

We thank the anonymous referee for their valuable comments which helped to improve this paper. This work has been supported in part by the National Astronomical Research Institute of Thailand (NARIT) and Srinakharinwirot University (grant no. 028/2564). SL acknowledges financial support from the Research Professional Development Project under the Science Achievement Scholarship of Thailand. The data in this work is based on observations obtained with XMM-Newton, an ESA science mission with instruments and contributions directly funded by ESA Member States and NASA.

\section*{Data availability}
The data underlying this article were accessed from {\it XMM-Newton} Observatory (\url{http://nxsa.esac.esa.int}). The derived data generated in this research will be shared on reasonable request to the corresponding author.




\bibliographystyle{mnras}
\bibliography{references} 








\bsp	
\label{lastpage}
\end{document}